\documentclass[10pt,a4paper,reqno]{amsart}
\usepackage{acronym}
\usepackage{amsfonts}
\usepackage{amsmath}
\usepackage{amssymb}
\usepackage{amsthm}
\usepackage{bm}
\usepackage{braket}
\usepackage{cite}
\usepackage{color}
\usepackage{geometry}
\usepackage{graphicx}
\usepackage{hyperref}
\usepackage{mathrsfs}
\usepackage{mathtools}
\usepackage{setspace}
\usepackage{stmaryrd}
\usepackage{subcaption}
\usepackage{tikz}

\usepackage{epsfig}
\usepackage{setspace}
\usepackage{booktabs}
\usepackage{threeparttable}
\usepackage{diagbox}
\usepackage{epstopdf}

\newgeometry{top=2cm,bottom=2cm,outer=1.5cm,inner=1.5cm}
\newcommand{\bse}{\begin{subequations}}
\newcommand{\ese}{\end{subequations}}

\numberwithin{equation}{section}

\title[PINN deep learning for solving rogue wave on the periodic background]{PINN deep learning for the Chen-Lee-Liu equation:
Rogue wave on the periodic background}
\author{Weiqi Peng}
\address[WP]{School of Mathematical Sciences, Shanghai Key Laboratory of Pure Mathematics and Mathematical Practice\\
East China Normal University \\ Shanghai 200062 \\ People's Republic of China}
\author{Juncai Pu}
\address[JP]{School of Mathematical Sciences, Shanghai Key Laboratory of Pure Mathematics and Mathematical Practice\\
East China Normal University \\ Shanghai 200062 \\ People's Republic of China}
\author{Yong Chen$^*$}
\address[YC]{School of Mathematical Sciences, Shanghai Key Laboratory of Pure Mathematics and Mathematical Practice\\
East China Normal University \\ Shanghai 200062 \\ People's Republic of China}
\address[YC]{College of Mathematics and Systems Science \\ Shandong University of Science and Technology \\ Qingdao 266590 \\ People's Republic of China}
\address[YC]{Department of Physics \\ Zhejiang Normal University \\ Jinhua 321004 \\ People's Republic of China}
\email{ychen@sei.ecnu.edu.cn(Corresponding author).}

\begin{document}

\begin{abstract}
We consider the exact rogue periodic wave (rogue  wave on the periodic background) and periodic wave solutions for the Chen-Lee-Liu equation via the odd-th order Darboux transformation. Then, the multi-layer physics-informed neural networks (PINNs) deep learning method is applied to research the data-driven rogue periodic wave, breather wave, soliton wave and periodic wave solutions of well-known Chen-Lee-Liu equation. Especially, the data-driven rogue periodic wave is learned for the first time to solve the partial differential equation. In addition, using image simulation, the relevant dynamical behaviors and error analysis for there solutions are presented.  The numerical results indicate that the rogue periodic wave, breather wave, soliton wave and periodic wave solutions for Chen-Lee-Liu equation can be generated well by PINNs deep learning method.
\end{abstract}

\maketitle

\section{Introduction}
The derivative-type  nonlinear Schr\"{o}dinger equation can be
considered as a appropriate  model to describe some nonlinear phenomena in plasma astrophysics \cite{Gao-chos7},  fluid dynamics \cite{Gao-chos9}, and nonlinear optics\cite{Gao-chos10,Gao-chos11}. The second type derivative nonlinear Schr\"{o}dinger(DNLSII) equation is  \cite{He-JMP5}
\begin{align}\label{0.02}
iq_{t}+q_{xx}+iqq^{\ast}q_{x}=0,
\end{align}
where the asterisk $\ast$ means the complex conjugation.
Eq.\eqref{0.02}  is usually named Chen-Lee-Liu(CLL) equation, which was first introduced by Chen et al.\cite{He-JMP5}.
The CLL equation is known as a model to simulate the propagation of the self-steepening optical pulses without
self-phase modulation\cite{He-LMP17}. Using the Hirota method,
the exact $N$-soliton solution of the CLL equation was constructed \cite{He-CLL2,He-CLL3}. The breather solution, rogue wave solution and rational soliton solution have been obtained based on the Darboux transformation(DT)\cite{Yang-CLL, He-CLL}. The initial-boundary value problem for the CLL equation was analysed on the half line via  the Fokas unified method\cite{Fan-CLL}.
There are other two type of derivative nonlinear Schr\"{o}dinger equations, including  first type derivative nonlinear Schr\"{o}dinger(DNLSI) equation and third type derivative nonlinear Schr\"{o}dinger(DNLSIII) equation. The DNLSI equation is \cite{He-JMP4}
\begin{align}\label{0.01}
q_{t}+iq_{xx}+(|q|^{2}q)_{x}=0.
\end{align}

The DNLSIII  takes the form \cite{He-JMP6}
\begin{align}\label{0.03}
iq_{t}+q_{xx}-iq^{2}q_{x}^{\ast}+\frac{1}{2}q^{3}(q^{\ast})^{2}=0,
\end{align}
Through the gauge transformations, the three kinds of
DNLS equations can be related to each other \cite{Fan-ZAMP3,Fan-ZAMP4}.
Eq.\eqref{0.01}, also called the Kaup-Newell (KN) equation, can be used to describe the behaviors
of small-amplitude Alfv\'{e}n waves in a low-$\beta$ plasma
\cite{He-RRP2,He-RRP3,He-RRP4}  and large-amplitude magnetohydrodynamic (MHD) waves in a high-$\beta$
plasma\cite{He-RRP5,He-RRP6}. In addition, the transmission of sub-picosecond pulses in single-mode fiber is described by Eq.\eqref{0.01} \cite{He-RRP7,He-RRP8}. Eq.\eqref{0.03}, known as Gerdjikov-Ivanov (GI) equation, was pioneered by Gerdjikov and Ivanov in Ref.\cite{He-JMP6}. As well as, since the Eq.\eqref{0.03} has certain higher-order nonlinear effects, it can be viewed as an extension of the nonlinear Schr\"{o}dinger(NLS) equation.

Rogue waves have been gradually reported in diverse fields, such as the deep
ocean \cite{Gao-chos23}, the nonlinear optics \cite{Gao-chos24} and Bose-Einstein
condensation \cite{Gao-chos25} and so on. Over the last few decades, rogue waves emerging on a plane wave background  have been studied a lot and great progress has been made\cite{Peng-EPL12,Peng-EPL13,Peng-EPL14,Peng-EPL15,Peng-EPL16,Peng-EPL17}. However, there are a great deal of work remains to be carried out for the rogue waves on the periodic background, which we call here as rogue periodic waves,  and rogue periodic waves are more general and practical than ones on a plane wave background\cite{Gao-chos28}. Therefore, more and more researchers have paid attention to rogue periodic waves for various integrable equations including the NLS equation, modified
Korteweg-de Vries equation, Hirota equation,  and sine-Gordon equation etc.\cite{Gao-chos29,Gao-chos30,Peng-wave,Zhanghaiq,gengxg}. However, to our knowledge, the rogue periodic waves for CLL equation \eqref{0.02} have not been studied. Thus, it is necessary and meaningful to study the rogue periodic waves for CLL equation. Without loss of generality, the construction of rogue periodic waves is usually associated with cumbersome Jacobian elliptic functions\cite{He-RRP37,Gao-chos29}, but in this paper, we will apply a direct way to construct rogue periodic waves according to the odd-th order DT of the CLL equation.

Mechanics learning with the neural network method \cite{PuChen23,PuChen24,PuChen25} has been widely applied in a variety of fields\cite{wangya12,wangya13}.  Especially, it plays a huge role in solving differential equations\cite{PuChen26}. Recently, the physics-informed neural network (PINN)\cite{PuChen28}  and its improvement \cite{PuChen29}  has been proposed to solve many linear and nonlinear differential equations. In general, based on PINN deep learning method, accurate solutions can be obtained with very small amounts of data. At the same time, since the underlying physical constraints are usually explicitly depicted by differential equations, the method also gives a better physical explanation for the predicted solution.
More recently, using PINN deep learning method,  Chen group constructed data-driven soliton solution for some nonlinear
evolution equations \cite{PuChen40,PuChen43,PuChen44} and data-driven high-order breather wave, rogue waves for the NLS equation and KN equation\cite{PuChenNLS,PuChenDNLS} with different initial and boundary conditions. Also, the data-driven rogue waves  were studied for the defocusing NLS equation with a potential\cite{Yan-wang} and high-order NLS equation\cite{daichaoq}. However, as far as we know, PINN deep learning for solving rogue periodic waves involving the partial differential equations has not been reported so far. Therefore, it will be very interesting and meaningful to research the data-driven rogue periodic wave via PINN deep learning method. As well as, the soliton wave, breather wave, periodic wave solutions  of the CLL equation have not been investigated by the PINN deep learning method. For all of these reasons, we will aim at solving the data-driven rogue periodic wave,  periodic wave, soliton wave and breather wave solutions for the CLL equation via deep learning.

The outline of this paper is organized as follows: In Sec.
2, the PINN deep leaning method is introduced for the general (1+1)-dimensional nonlinear integrable systems. In Sec. 3, we derive the exact periodic wave solution and rogue periodic wave solution for the CLL equation \eqref{0.02} in terms of the odd-th order Darboux transformation (DT). In Sec. 4, by applying the PINN deep learning approach, the data-driven periodic wave, rogue periodic wave, soliton wave and breather wave solutions of  the CLL equation \eqref{0.02} are investigated. In Sec. 5, we give some conclusions and discussions.

\section{The PINN deep learning method}
The (1+1)-dimensional complex nonlinear dispersive equations in its general form can be  written as
\begin{align}\label{0.1}
q_{t}+\mathcal{N}_{q}(q, q_{x}, q_{xx}, q_{xxx},\cdots)=0,
\end{align}
where $q$ is a complex valued function with variables $x$ and $t$. and $\mathcal{N}_{q}$ is some
nonlinear function of the $q$ and its derivatives of arbitrary orders with respect to $x$. Taking $q=u+iv$,  we decompose the above complex equation \eqref{0.1} into following  two real nonlinear dispersive equations, given by
\begin{align}\label{0.2}
u_{t}+\mathcal{N}_{u}(u, u_{x}, u_{xx}, u_{xxx},\cdots)=0,
\end{align}
\begin{align}\label{0.3}
v_{t}+\mathcal{N}_{v}(v, v_{x}, v_{xx}, v_{xxx},\cdots)=0.
\end{align}

Then the physics-informed neural networks $f_{u}(x, t)$ and $f_{v}(x, t)$ can be defined as
\begin{align}\label{0.4}
f_{u}:=u_{t}+\mathcal{N}_{u}(u, u_{x}, u_{xx}, u_{xxx},\cdots),
\end{align}
\begin{align}\label{0.5}
f_{v}:=v_{t}+\mathcal{N}_{v}(v, v_{x}, v_{xx}, v_{xxx},\cdots),
\end{align}
where $\mathcal{N}_{u}(u, u_{x}, u_{xx}, u_{xxx},\cdots), \mathcal{N}_{v}(v, v_{x}, v_{xx}, v_{xxx},\cdots)$ are the  physical models given in Eq.\eqref{0.2}, \eqref{0.3}, and  $u(x, t; w, b)$, $v(x, t; w, b)$ are the latent function of the deep neural network with the weight parameter $w$ and bias parameter $b$, which can be used to approximate the exact complex-valued solution $q(x, t)$ of objective equations. Then the networks $f_{u}(x, t), f_{v}(x, t)$ can also be found with the help of automatic differentiation mechanism in deep learning \cite{PuChen420}.
By using the multi-hidden-layer deep neural network, the network parameters of the latent functions $u, v$ and  networks $f_{u}(x, t)$ and $f_{v}(x, t)$ can be constantly trained.

Throughout the training process, in order to obtain the optimum training results, we use L-BFGS optimization method \cite{wangya44} to minimize the whole mean squared error, that is, the loss function
\begin{align}\label{0.6}
Loss_{\Theta}=Loss_{u}+Loss_{v}+Loss_{f_{u}}+Loss_{f_{v}},
\end{align}
where
\begin{align}\label{0.7}
Loss_{u}=\frac{1}{N_{q}}\sum_{i=1}^{N_{q}}|u(x_{q}^{i},t_{q}^{i})-u^{i}|^{2},\quad Loss_{v}=\frac{1}{N_{q}}\sum_{i=1}^{N_{q}}|v(x_{q}^{i},t_{q}^{i})-v^{i}|^{2},
\end{align}
and
\begin{align}\label{0.8}
Loss_{f_{u}}=\frac{1}{N_{f}}\sum_{j=1}^{N_{f}}|f_{u}(x_{f}^{j},t_{f}^{j})|^{2},\quad
Loss_{f_{v}}=\frac{1}{N_{f}}\sum_{j=1}^{N_{f}}|f_{v}(x_{f}^{j},t_{f}^{j})|^{2},
\end{align}
where $\{x_{q}^{i}, t_{q}^{i}, u^{i}\}_{i=1}^{N_{q}}$ and $\{x_{q}^{i}, t_{q}^{i}, v^{i}\}_{i=1}^{N_{q}}$ are the sampled initial and boundary value  training data of $q(x, t)$. Similarly, the collocation
points for $f_{u}(x, t)$ and $f_{v}(x, t)$ are marked by $\{x_{f}^{j}, t_{f}^{j}\}_{j=1}^{N_{f}}$ and $\{x_{f}^{j}, t_{f}^{j}\}_{j=1}^{N_{f}}$. The loss function \eqref{0.6} contains the  losses of
initial-boundary value data and the losses of  networks \eqref{0.4} and \eqref{0.5} at a finite set of collocation points.
Of which, the first two items on the right hand side of Eq.\eqref{0.6} attempt to let the learning solution  approaches the exact one for the the initial and boundary value data, and the latter two on the right hand side make the hidden $u, v$ satisfy the target nonlinear dispersive equation \eqref{0.2}, \eqref{0.3}.

In this paper, the simple multilayer perceptrons (i.e., feedforward neural networks) with the Xavier initialization are chosen as  the neural network model, and we select the hyperbolic tangent (tanh) as  activation function. All codes are based on Python 3.7 and Tensorflow 1.15, and all numerical experiments shown here are run on a DELL Precision 7920 Tower computer with 2.10 GHz 8-core Xeon Silver 4110 processor and 64-GB memory.

\section{The exact periodic wave and rogue periodic waves}
In this section, we are committed to given the exact periodic wave and rogue periodic waves solutions for the CLL  equation \eqref{0.02} via DT. The CLL equation\eqref{0.02} is associated with the following spectral problem
\begin{gather}
\Phi_{x}=U\Phi=(-i\lambda^{2}-\frac{i}{4}qr)\sigma_{3}\Phi+\lambda Q\Phi,\notag\\
\Phi_{t}=V\Phi=(-2i\lambda^{4}-iqr\lambda^{2}-\frac{1}{4}(qr_{x}-rq_{x})-\frac{i}{8}q^{2}r^{2})\sigma_{3}\Phi+2\lambda^{3}Q\Phi+\lambda P\Phi,\label{A1}
\end{gather}
with
\begin{gather}
\Phi(x, t, \lambda)=\left(\begin{array}{c}
     \phi(x, t, \lambda)\\
  \varphi(x, t, \lambda)\\
\end{array}\right),\ \sigma_{3}=\left(\begin{array}{cc}
     1 &  0\\
     0&  -1\\
\end{array}\right),\ Q=\left(\begin{array}{cc}
     0 &  q\\
     r&  0\\
\end{array}\right),\notag\\
P=\left(\begin{array}{cc}
     0  &  iq_{x}+\frac{1}{2}q^{2}r\\
     -ir_{x}+\frac{1}{2}r^{2}q  &  0\\
\end{array}\right).\label{A2}
\end{gather}
Under the reduction condition $r=-q^{\ast}$, the CLL equation \eqref{0.02} can be raised by the compatibility condition \eqref{A1}. Moreover, to keep the above reduction condition invariant after each step DT, the Lax pair equations should meet following symmetry conditions as\\
(1). $\lambda_{k}=-\lambda_{k}^{\ast}$, $\phi^{\ast}_{k}(x, t, \lambda_{k})=\varphi_{k}(x, t, \lambda_{k})$;\\
(2). $\lambda_{2k}=-\lambda_{2k-1}^{\ast}$, $\phi^{\ast}_{2k-1}(x, t, \lambda_{2k-1})=\varphi_{2k}(x, t, \lambda_{2k})$, $\varphi^{\ast}_{2k-1}(x, t, \lambda_{2k-1})=\phi_{2k}(x, t, \lambda_{2k})$.\\

Let $\Phi_{k}(x, t, \lambda_{k})=[\phi_{k}(x, t, \lambda_{k}), \varphi_{k}(x, t, \lambda_{k})]^{T}$ with $T$ being the
matrix transpose are the distinct solutions of Lax Pair \eqref{A1} related to $\lambda_{k}$, and the seed solution is
$q^{[0]}=Ae^{i\theta}, \theta=ax-(aA^{2}+a^{2})t$, of which $a$ and $A$ being the complex parameters, then the $N$-th order analytic solutions for CLL equation \eqref{0.02} are written into the following determinant expression\cite{Yang-CLL,He-CLL}
\begin{align}\label{A3}
q^{[N]}=e^{i\eta(\frac{1+(-1)^{N+1}}{2})}\left(\frac{q^{[0]}\det (S)+2i\det(W)}{\det (S^{\ast})}\right),\ e^{i\eta}=e^{\frac{iA^{2}x}{2}-(iaA^{2}+\frac{iA^{4}}{4})t},
\end{align}
with $W=(W_{1},W_{2},\cdots, W_{N}), S=(S_{1},S_{2},\cdots, S_{N})$, and\\
(i) $N=2n+1$,
\begin{gather}
W_{k}=(\varphi_{k}, \lambda_{k}\phi_{k}, \cdots, \lambda_{k}^{2n-2}\varphi_{k}, \lambda^{2n-1}_{k}\phi_{k}, -\lambda^{2n+1}_{k}\phi_{k})^{T},\notag\\
S_{k}=(\varphi_{k}, \lambda_{k}\phi_{k},\cdots,  \lambda_{k}^{2n-2}\varphi_{k}, \lambda^{2n-1}_{k}\phi_{k}, \lambda^{2n}_{k}\varphi_{k})^{T}.\label{A4}
\end{gather}
(ii) $N=2n$,
\begin{align}\label{A5}
&W_{k}=(\phi_{k}, \lambda_{k}\varphi_{k},\cdots,  \lambda_{k}^{2n-3}\varphi_{k}, \lambda^{2n-2}_{k}\phi_{k}, -\lambda^{2n}_{k}\phi_{k})^{T},\notag\\
&S_{k}=(\phi_{k}, \lambda_{k}\varphi_{k}, \cdots,  \lambda_{k}^{2n-3}\varphi_{k},  \lambda^{2n-2}_{k}\phi_{k}, \lambda^{2n-1}_{k}\varphi_{k})^{T}.
\end{align}

Let $q^{[0]}=Ae^{i\theta}, \theta=ax-(aA^{2}+a^{2})t$ become the seed solution, and solving the
Lax pair equation \eqref{A1}, we can obtain the corresponding vector eigenfunctions $\Phi_{k}$ associated with $\lambda_{k}$, given by
\begin{align}\label{4.1}
\Phi_{k}(\lambda_{k})=\left(\begin{array}{c}
     \phi_{k}(x, t, \lambda_{k})\\
  \varphi_{k}(x, t, \lambda_{k})\\
\end{array}\right)=\left(\begin{array}{c}
      \psi_{1}(\lambda_{k})+\psi^{\ast}_{2}(-\lambda_{k}^{\ast})\\
      \psi_{2}(\lambda_{k})+\psi^{\ast}_{1}(-\lambda_{k}^{\ast}) \\
\end{array}\right),
\end{align}
with

\begin{align}\label{4.2}
\left(\begin{array}{c}
      \psi_{1}(\lambda_{k})\\
      \psi_{2}(\lambda_{k})\\
\end{array}\right)=\left(\begin{array}{c}
      \frac{-i(A^{2}-4\lambda_{k}^{2}-2a+s)}{4\lambda_{k}}e^{\frac{i}{2}(\frac{s}{2}x+bt+\theta)}\\
      Ae^{\frac{i}{2}(\frac{s}{2}x+bt-\theta)}\\
\end{array}\right),
\end{align}
where
\begin{gather}
s=\sqrt{A^{4}+8\lambda_{k}^{2}A^{2}+16\lambda_{k}^{4}-4A^{2}a+16a\lambda_{k}^{2}+4a^{2}},\notag\\
b=\frac{A^{2}}{4}-2A^{2}\lambda_{k}^{2}+4\lambda_{k}^{4}-\frac{A^{2}}{4}(A^{2}-4\lambda_{k}^{2}-2a+s)\notag\\
+\lambda_{k}^{2}(A^{2}-4\lambda_{k}^{2}-2a+s)-\frac{a}{2}(A^{2}-4\lambda_{k}^{2}-2a+s)-a^{2}.\label{4.3}
\end{gather}

For CLL equation \eqref{0.02}, as presented in Ref. \cite{Yang-CLL,He-CLL}, the rational solution, breather wave and rogue wave on the constant background have been constructed via  expression \eqref{A3} when $N=2n$.
However, we hereby try to  construct the  periodic wave and
rogue periodic wave for CLL equation by taking  $N=2n+1$ in expression \eqref{A3}.

Taking $N=1$, and $A=1, a=-1$  in Eq.\eqref{A3}, the exact one periodic wave solution is derived as
\begin{align}\label{5}
q(x, t)=e^{\frac{i}{2}x+\frac{3i}{4}t}\left(\frac{q^{[0]}\varphi_{1}(\lambda_{1})-2i\lambda_{1}\phi_{1}(\lambda_{1})}
{\varphi^{\ast}_{1}(\lambda_{1})}\right)=\frac{M_{1}}{N_{1}}
\end{align}
with
\begin{gather}
M_{1}=(8\beta^{3}+2c\beta+2\beta)e^{-\frac{1}{8}ic(4t\beta^{2}-t-2x)+\frac{3it}{4}}+(-4\beta^{2}+c+3)e^{\frac{1}{8}ic(4t\beta^{2}-t-2x)+\frac{3it}{4}},\notag\\
N_{1}=(3+4\beta^{2}+c)e^{-\frac{1}{8}ic(4t\beta^{2}-t-2x)-\frac{ix}{2}}-4\beta e^{\frac{1}{8}ic(4t\beta^{2}-t-2x)-\frac{ix}{2}},\label{6}
\end{gather}
where $c=\sqrt{16\beta^{4}+8\beta^{2}+9}$, and $\beta$ is the arbitrary constant. The dynamic behaviors of exact one periodic wave solution are shown in Fig. 1, and it is obvious that the solution \eqref{5} is a periodic solution  in the $x$ direction.
\\
\\
{\rotatebox{0}{\includegraphics[width=4.3cm,height=3.6cm,angle=0]{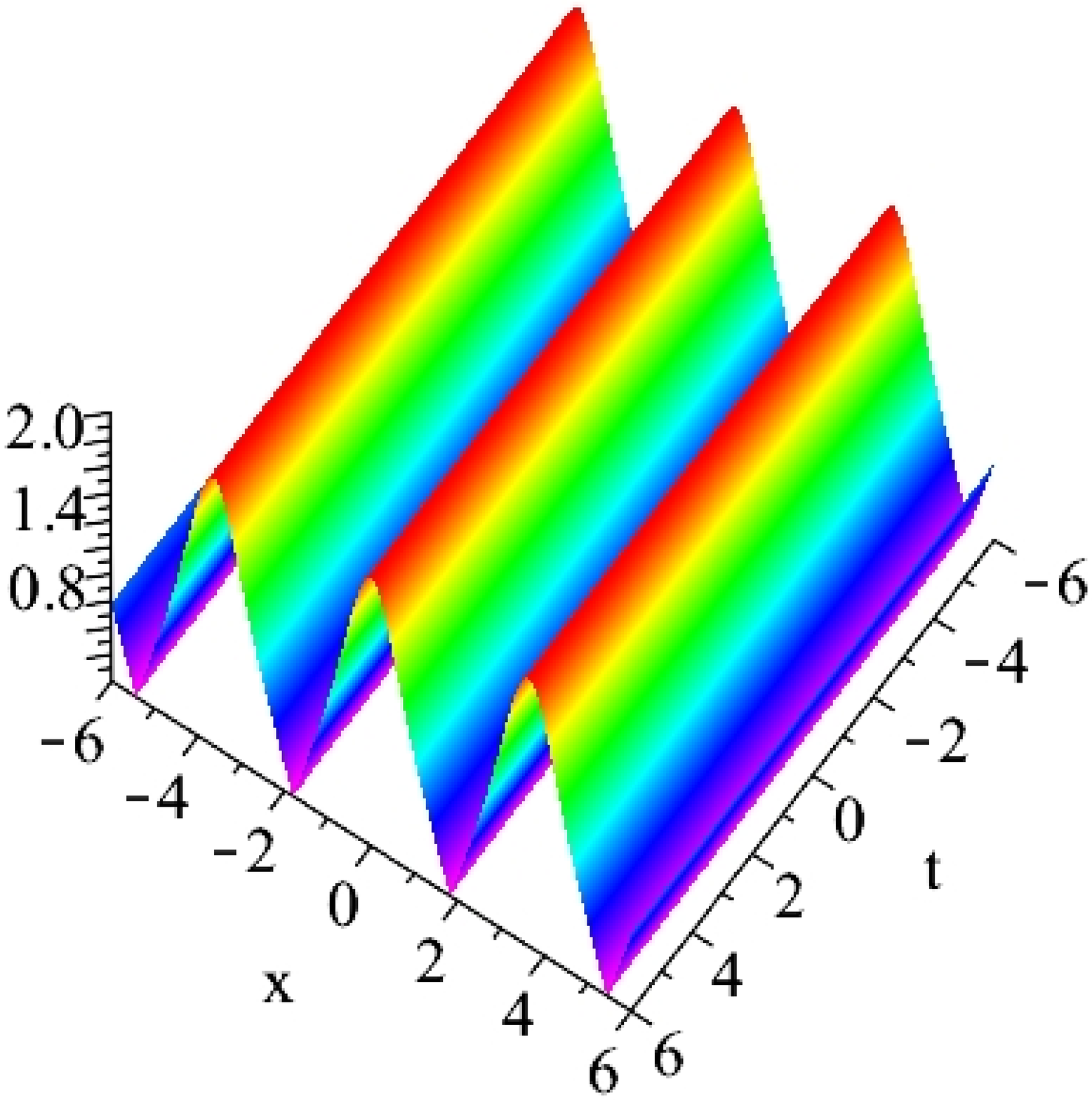}}}
~~~~
{\rotatebox{0}{\includegraphics[width=4.3cm,height=3.6cm,angle=0]{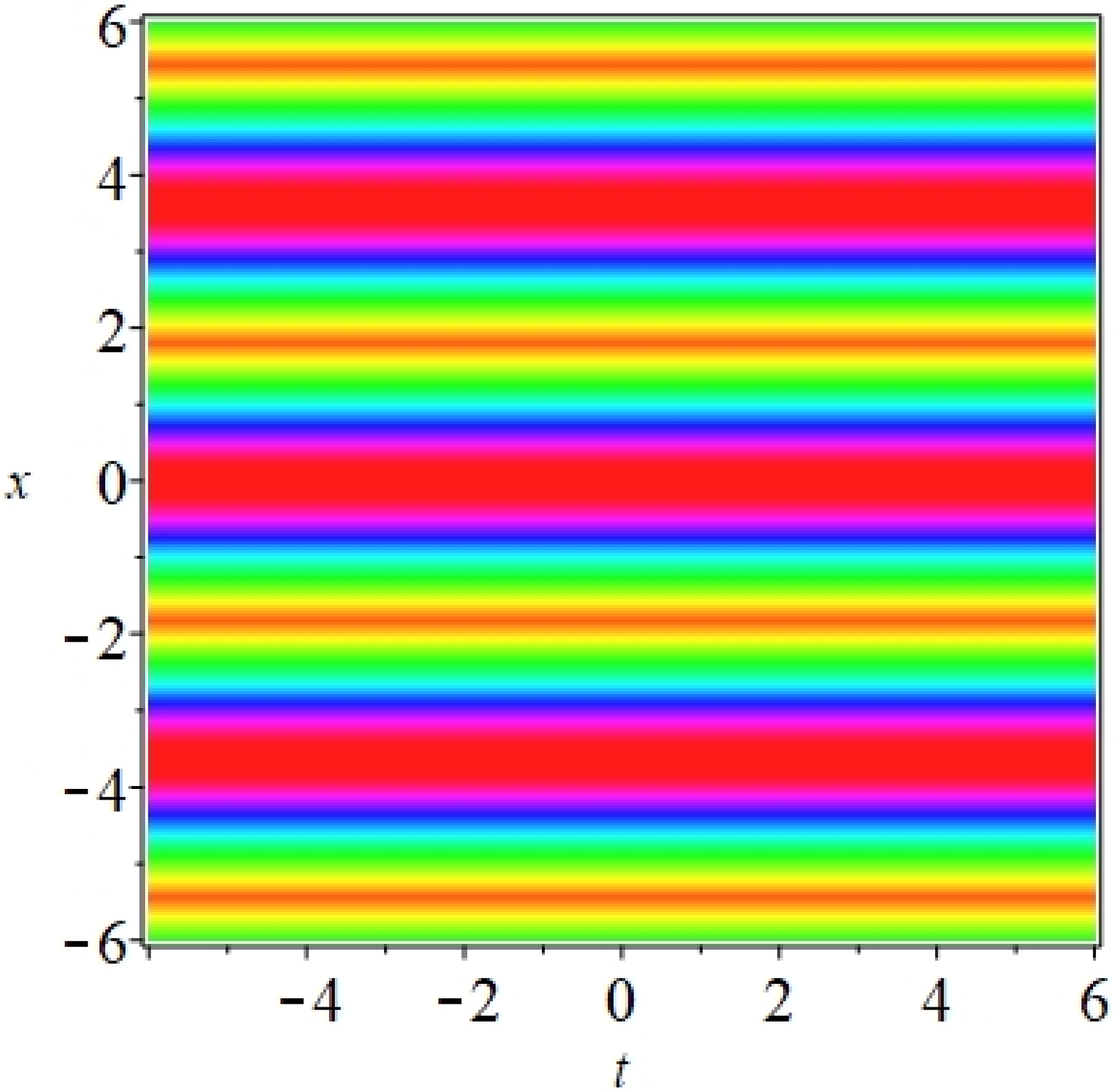}}}
~~~~
{\rotatebox{0}{\includegraphics[width=4.3cm,height=3.6cm,angle=0]{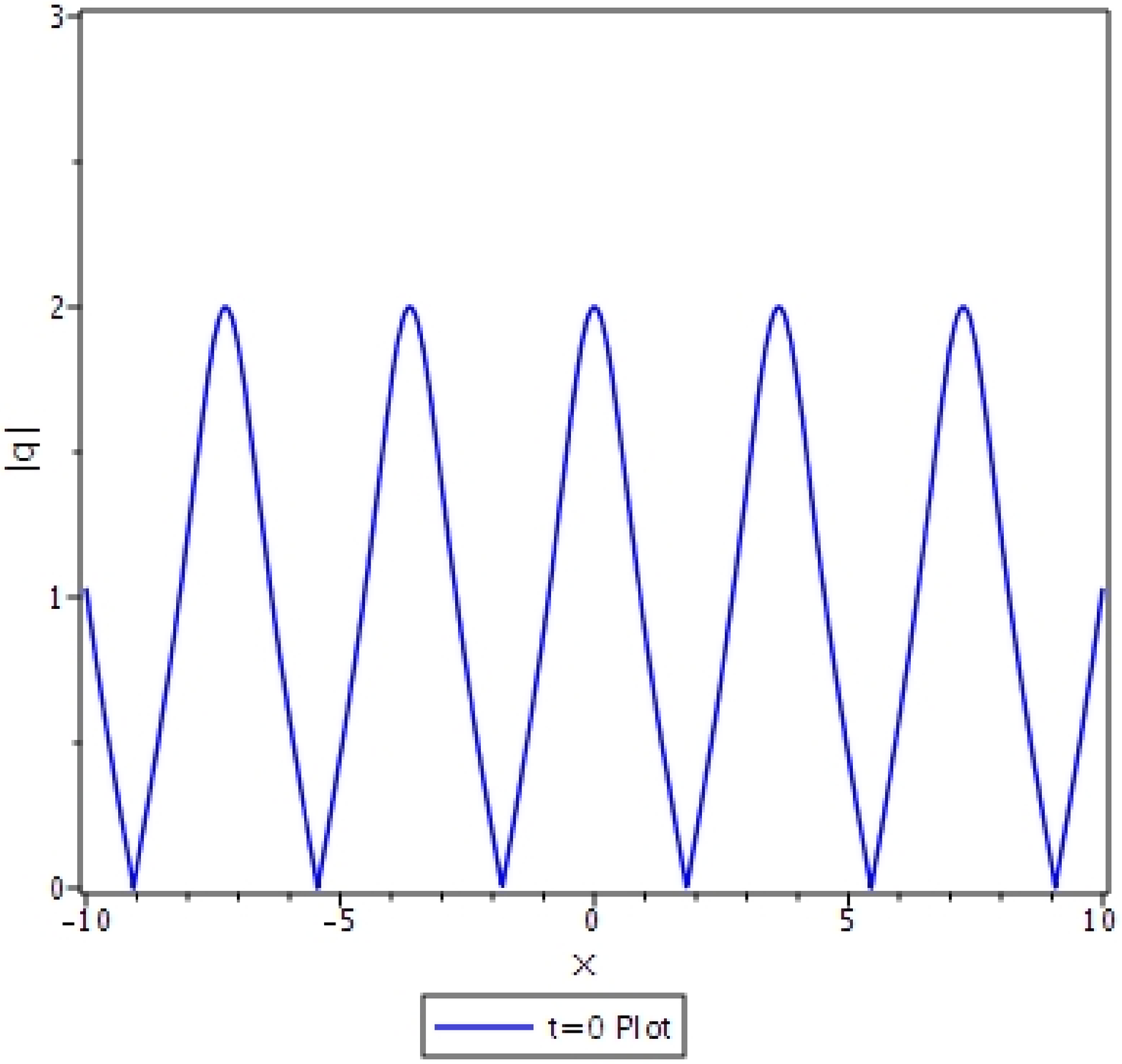}}}\\

$\quad\qquad\textbf{(a)}\qquad\qquad\qquad\qquad\qquad\quad\qquad
 \textbf{(b)}\qquad\qquad\qquad\qquad\qquad\quad\textbf{(c)}$\\
\noindent { \small \textbf{Figure 1.} (Color online) The one periodic solution for Eq.\eqref{0.02} with parameter $\beta=0.5$. $\textbf{(a)}$ Three dimensional plot;
$\textbf{(b)}$ The density plot;
$\textbf{(c)}$ The wave propagation along the $x$-axis at $t=0$.}\\

Let $N=2n+1$, $\lambda_{1}=\frac{\sqrt{-2a}}{2}+\frac{i}{2}A, \lambda_{2}=-\lambda_{1}^{\ast}$ and $\lambda_{N}=i\beta$. By taking Taylor expansion as \cite{He-KN12,Gao-GI} in \eqref{A3} for $N=2n+1$, the exact $n$-th rogue periodic wave solution can be given by
\begin{align}\label{6.1}
q^{[n]}=e^{i\eta}\left(\frac{q^{[0]}\det (\tilde{S})+2i\det(\tilde{W})}{\det (\tilde{S}^{\ast})}\right),
\end{align}
with
\begin{gather}
\tilde{W}=\left(\begin{array}{cccc}
     \varphi[1,0,1]  &  \varphi[2,0,1] &  \varphi[1,0,2] &  \varphi[2,0,2] \\
     \phi[1,1,1]  &  \phi[2,1,1] &  \phi[1,1,2] &  \phi[2,1,2] \\
     \vdots  &  \vdots &  \vdots &  \vdots \\
     \varphi[1,N-3,1]  &  \varphi[2,N-3,1] &  \varphi[1,N-3,2] &  \varphi[2,N-3,2] \\
     \phi[1,N-2,1]  &  \phi[2,N-2,1] &  \phi[1,N-2,2] &  \phi[2,N-2,2] \\
     -\phi[1,N,1]  &  -\phi[2,N,1] &  -\phi[1,N,2] &  -\phi[2,N,2] \\
\end{array}\right.,\notag\\
\left.\begin{array}{cccc}
       \cdots &  \varphi[1,0,n] &  \varphi[2,0,n] &  \varphi_{N}\\
       \cdots &  \phi[1,1,n] &  \phi[2,1,n] &  \lambda_{N}\phi_{N}\\
      \cdots &  \vdots &  \vdots &  \vdots\\
       \cdots &  \varphi[1,N-3,n] &  \varphi[2,N-3,n] &  \lambda_{N}^{N-3}\varphi_{N}\\
     \cdots &  \phi[1,N-2,n] &  \phi[2,N-2,n] &  \lambda_{N}^{N-2}\phi_{N}\\
     \cdots &  -\phi[1,N,n] &  -\phi[2,N,n] &  -\lambda_{N}^{N}\phi_{N}\\
\end{array}\right),
\end{gather}

\begin{gather}
\tilde{S}=\left(\begin{array}{cccccccc}
     \varphi[1,0,1]  &  \varphi[2,0,1] &  \varphi[1,0,2] &  \varphi[2,0,2] \\
     \phi[1,1,1]  &  \phi[2,1,1] &  \phi[1,1,2] &  \phi[2,1,2]\\
     \vdots  &  \vdots &  \vdots &  \vdots \\
     \varphi[1,N-3,1]  &  \varphi[2,N-3,1] &  \varphi[1,N-3,2] &  \varphi[2,N-3,2] \\
     \phi[1,N-2,1]  &  \phi[2,N-2,1] &  \phi[1,N-2,2] &  \phi[2,N-2,2] \\
     \varphi[1,N-1,1]  &  \varphi[2,N-1,1] &  \varphi[1,N-1,2] &  \varphi[2,N-1,2] \\
\end{array}\right.,\notag\\
\left.\begin{array}{cccccccc}
       \cdots &  \varphi[1,0,n] &  \varphi[2,0,n] &  \varphi_{N}\\
       \cdots &  \phi[1,1,n] &  \phi[2,1,n] &  \lambda_{N}\phi_{N}\\
       \cdots &  \vdots &  \vdots &  \vdots\\
       \cdots &  \varphi[1,N-3,n] &  \varphi[2,N-3,n] &  \lambda_{N}^{N-3}\varphi_{N}\\
       \cdots &  \phi[1,N-2,n] &  \phi[2,N-2,n] &  \lambda_{N}^{N-2}\phi_{N}\\
      \cdots &  \varphi[1,N-1,n] &  \varphi[2,N-1,n] &  \lambda_{N}^{N-1}\varphi_{N}\\
\end{array}\right),\label{6.2}
\end{gather}
where
\begin{align}\label{6.3}
\phi[l,j,n]=\frac{1}{n!}\frac{\partial^{2n}}{\partial \epsilon^{2n}}[(\lambda_{l}+\epsilon^{2})^{j}\phi(\lambda_{l}+\epsilon^{2})], \ \varphi[l,j,n]=\frac{1}{n!}\frac{\partial^{2n}}{\partial \epsilon^{2n}}[(\lambda_{l}+\epsilon^{2})^{j}\varphi(\lambda_{l}+\epsilon^{2})].
\end{align}

Taking $N=3$  and $A=1, a=-1$ in Eq.\eqref{6.1} and using Maple symbolic computation, the exact one rogue periodic wave solution can be given by
\begin{align}\label{7}
q(x, t)=\frac{M_{3}}{D_{3}},
\end{align}
with
\begin{gather}
M_{3}=-96\beta M_{3}^{+}e^{-\frac{1}{8}ic(4t\beta^{2}-t-2x)+\frac{3it}{4}}-48M_{3}^{-}e^{\frac{1}{8}ic(4t\beta^{2}-t-2x)+\frac{3it}{4}},\notag\\
D_{3}=D_{3}^{+}e^{-\frac{1}{8}ic(4t\beta^{2}-t-2x)-\frac{ix}{2}}+96\beta D_{3}^{-}e^{\frac{1}{8}ic(4t\beta^{2}-t-2x)-\frac{ix}{2}},\notag\\
M_{3}^{+}=c[(\frac{x^{2}}{4}+(\frac{t}{2}+\frac{i}{6})x+\frac{1}{12}+\frac{3t^{2}}{4})\beta^{2}+\frac{x^{2}}{16}+(\frac{t}{8}+\frac{7i}{24})x-\frac{1}{6}+\frac{3t^{2}}{16}
+\frac{it}{4}]\notag\\
+(x^{2}+(2t+\frac{2i}{3})x+\frac{1}{3}+3t^{2})\beta^{4}+(\frac{x^{2}}{2}+(t+\frac{4i}{3})x-\frac{1}{6}+\frac{3t^{2}}{2}+it)\beta^{2}\notag\\
+\frac{9x^{2}}{16}
+(\frac{5i}{8}+\frac{9t}{8})x+\frac{27t^{2}}{16}-\frac{3it}{4}-\frac{9}{16},\notag\\
M_{3}^{-}=c[(\frac{x^{2}}{4}+(\frac{t}{2}+\frac{i}{6})x+\frac{1}{12}+\frac{3t^{2}}{4})\beta^{2}-\frac{3x^{2}}{16}+(-\frac{3t}{8}+\frac{i}{8})x+\frac{3}{16}-\frac{9t^{2}}{16}
+\frac{3it}{4}]\notag\\
-(x^{2}+(2t+\frac{2i}{3})x+\frac{1}{3}+3t^{2})\beta^{4}+(-\frac{x^{2}}{2}-(t+\frac{4i}{3})x+\frac{3}{2}-\frac{3t^{2}}{2}+it)\beta^{2}\notag\\
-\frac{9x^{2}}{16}
+(\frac{3i}{8}-\frac{9t}{8})x-\frac{27t^{2}}{16}+\frac{9it}{4}+\frac{9}{16},\notag\\
D_{3}^{+}=c[(-12x^{2}+(8i-24t)x-36t^{2}-4)\beta^{2}-9x^{2}-(6i+18t)x-27t^{2}-3]\notag\\
+(-48x^{2}+(32i-96t)x-144t^{2}-16)\beta^{4}+(-24x^{2}+(-32i-48t)x\notag\\
-96it-72t^{2}-40)\beta^{2}
-27x^{2}-(18i+54t)x-81t^{2}-9,\notag\\
D_{3}^{-}=c[-\frac{x^{2}}{8}+(\frac{i}{12}-\frac{t}{4})x+\frac{it}{4}+\frac{1}{24}-\frac{3t^{2}}{8}]
+(it+\frac{1}{3})\beta^{2}+\frac{3it}{4}+\frac{ix}{2}+\frac{1}{4}.\label{8}
\end{gather}

The dynamic behaviors of exact one rogue periodic wave solution \eqref{7} are shown in Fig. 2. From Fig. 2, it is easily to find  that there is a rogue wave that arises in a background of periodic wave, and the rogue wave is distributed in the region where the periodic  wave reaches its amplitude.
\\
\\
{\rotatebox{0}{\includegraphics[width=4.3cm,height=3.6cm,angle=0]{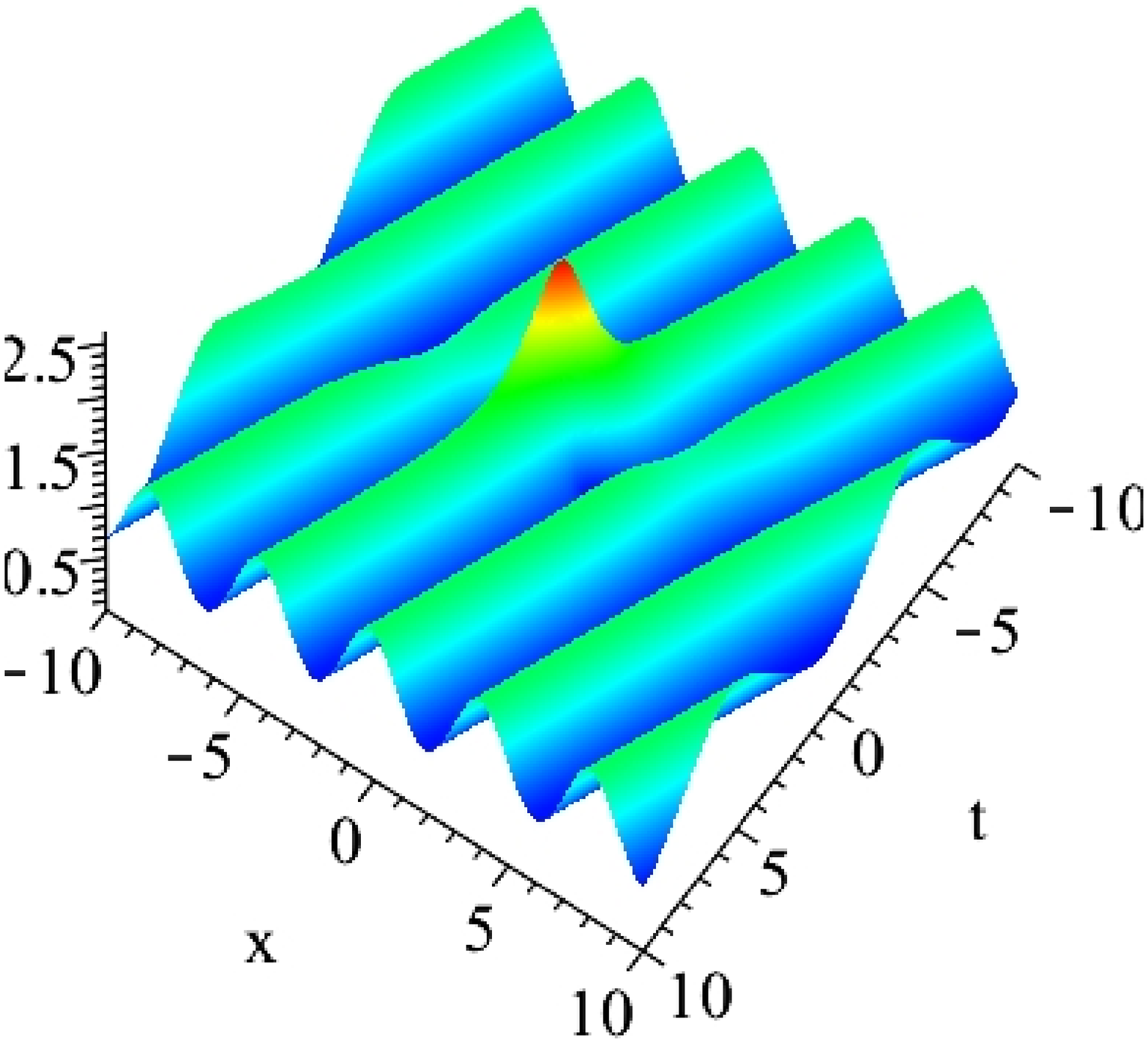}}}
~~~~
{\rotatebox{0}{\includegraphics[width=4.3cm,height=3.6cm,angle=0]{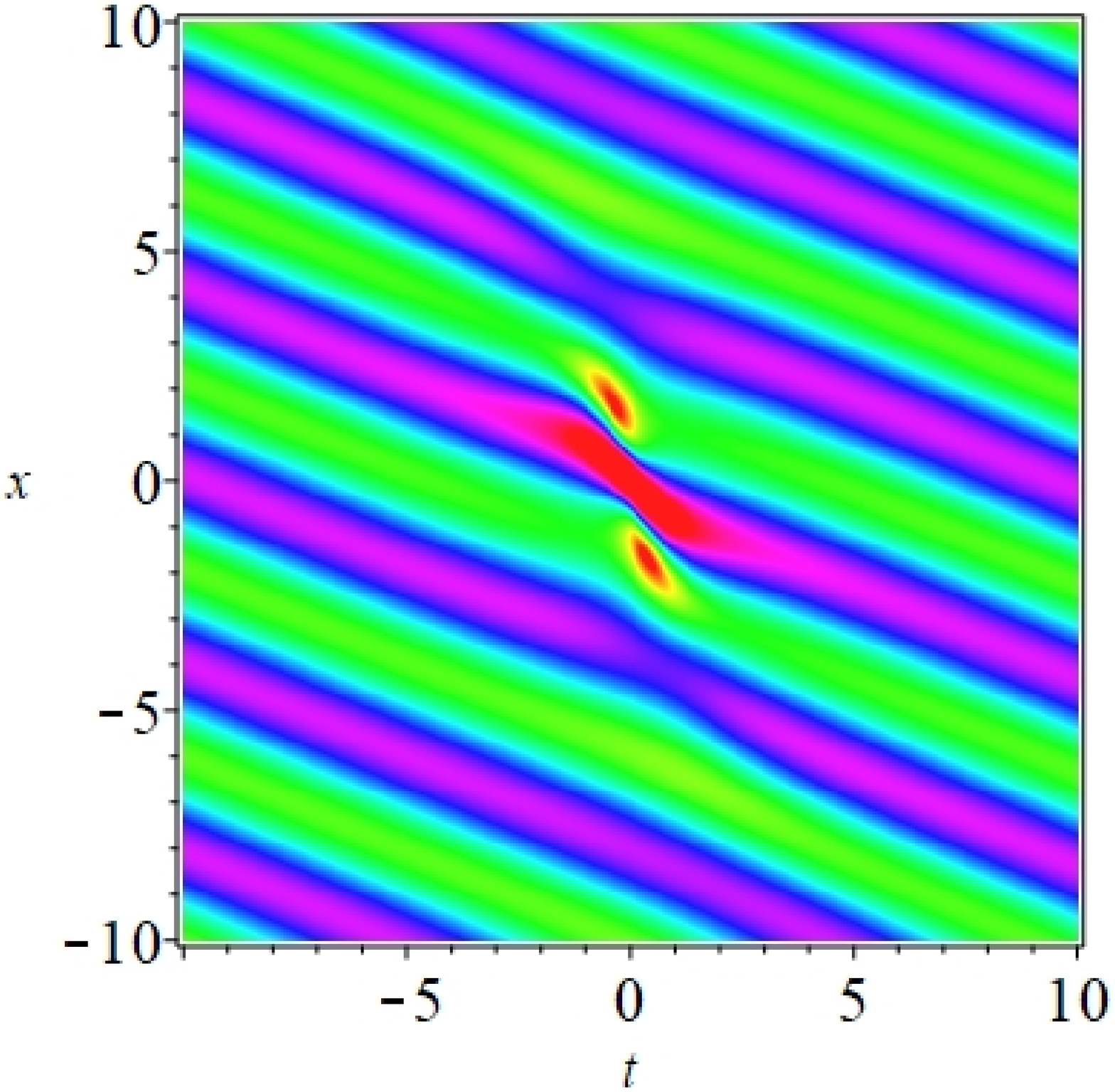}}}
~~~~
{\rotatebox{0}{\includegraphics[width=4.3cm,height=3.6cm,angle=0]{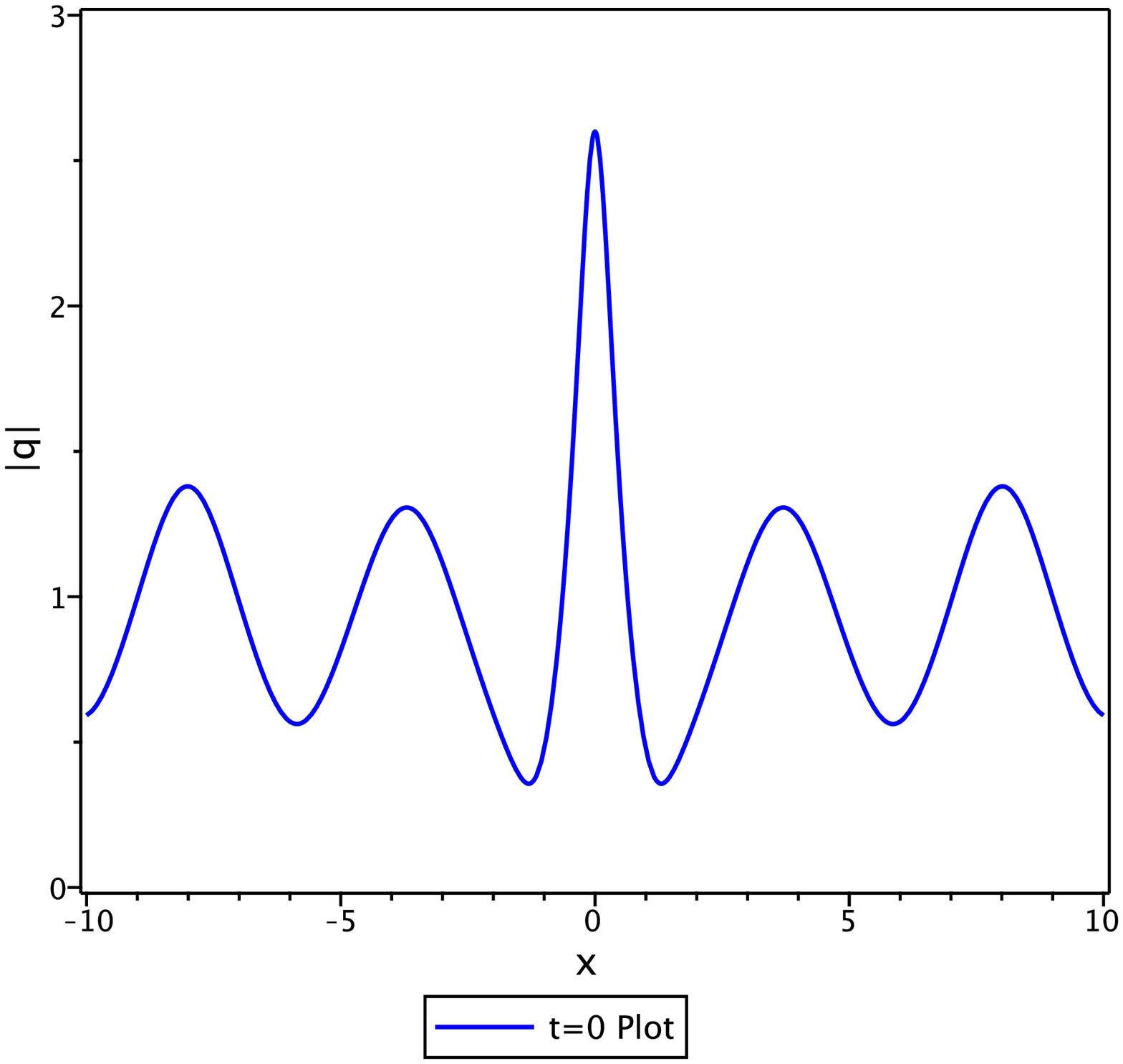}}}\\

$\quad\qquad\textbf{(a)}\qquad\qquad\qquad\qquad\qquad\quad\qquad
 \textbf{(b)}\qquad\qquad\qquad\qquad\qquad\quad\textbf{(c)}$\\
\noindent { \small \textbf{Figure 2.} (Color online) The one rogue periodic wave solution \eqref{7} for Eq.\eqref{0.02} with parameter $\beta=-0.2$. $\textbf{(a)}$ Three dimensional plot;
$\textbf{(b)}$ The density plot;
$\textbf{(c)}$ The wave propagation along the $x$-axis at $t=0$.}\\
\\
\\
{\rotatebox{0}{\includegraphics[width=4.3cm,height=3.6cm,angle=0]{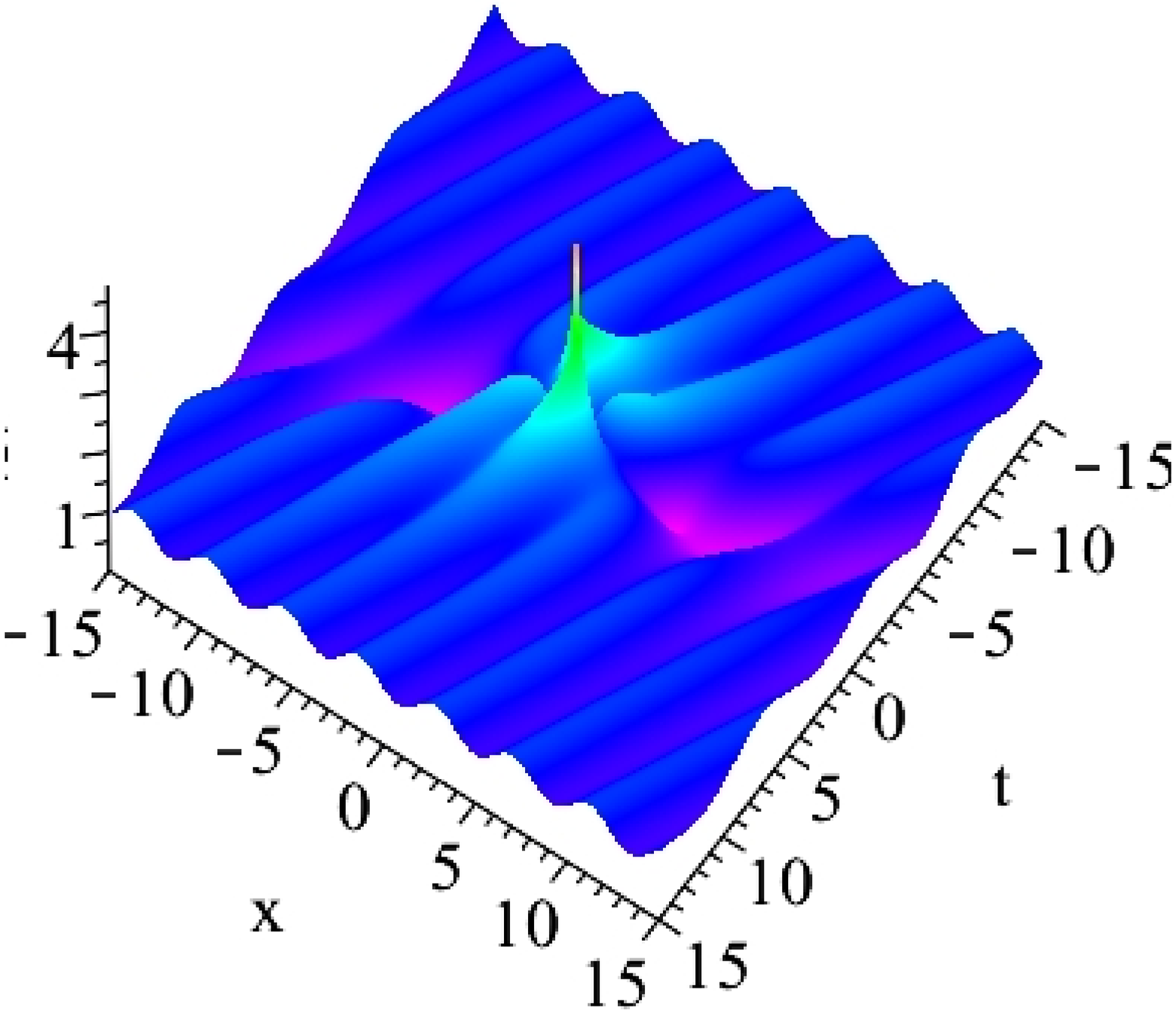}}}
~~~~
{\rotatebox{0}{\includegraphics[width=4.3cm,height=3.6cm,angle=0]{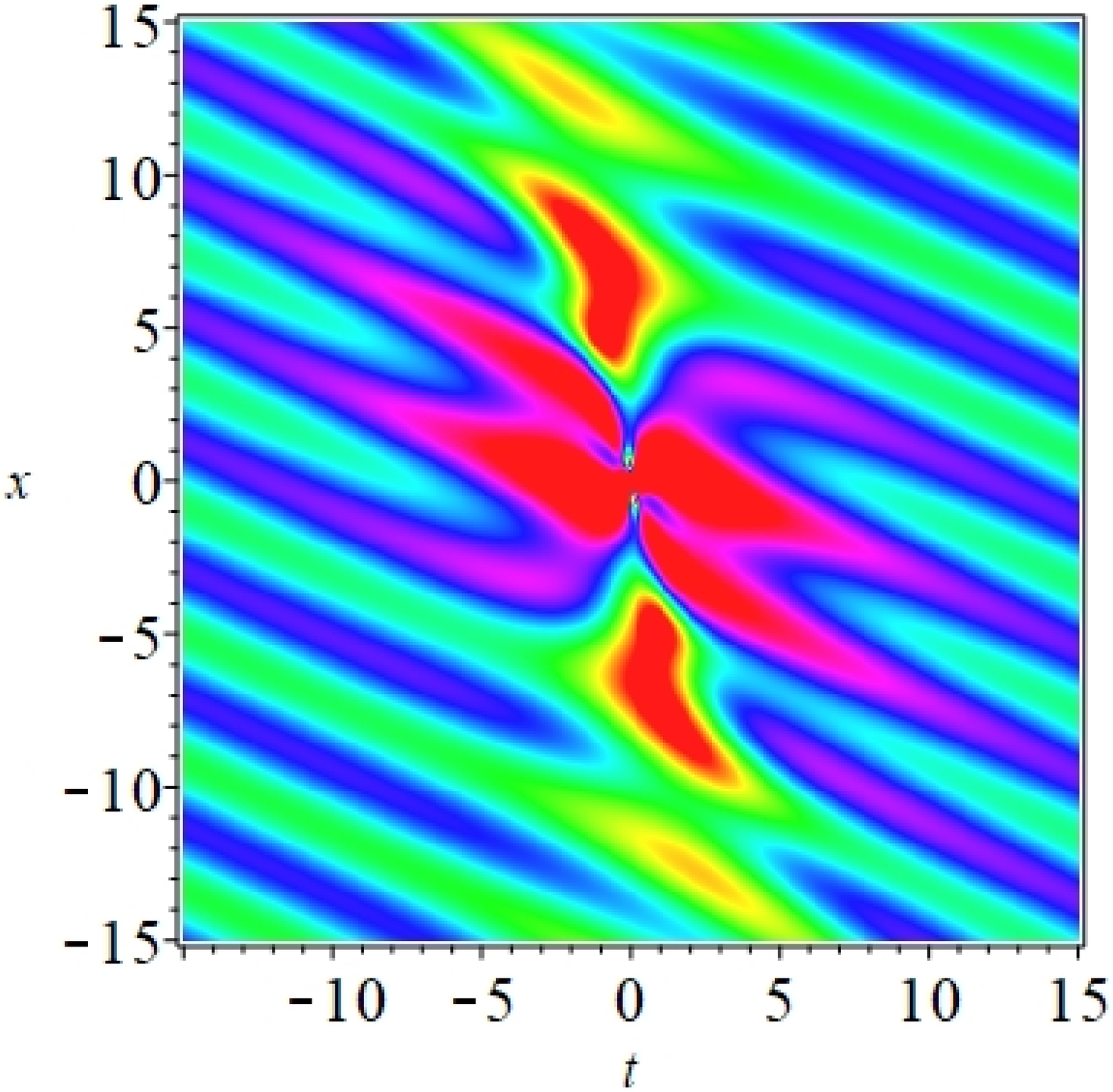}}}
~~~~
{\rotatebox{0}{\includegraphics[width=4.3cm,height=3.6cm,angle=0]{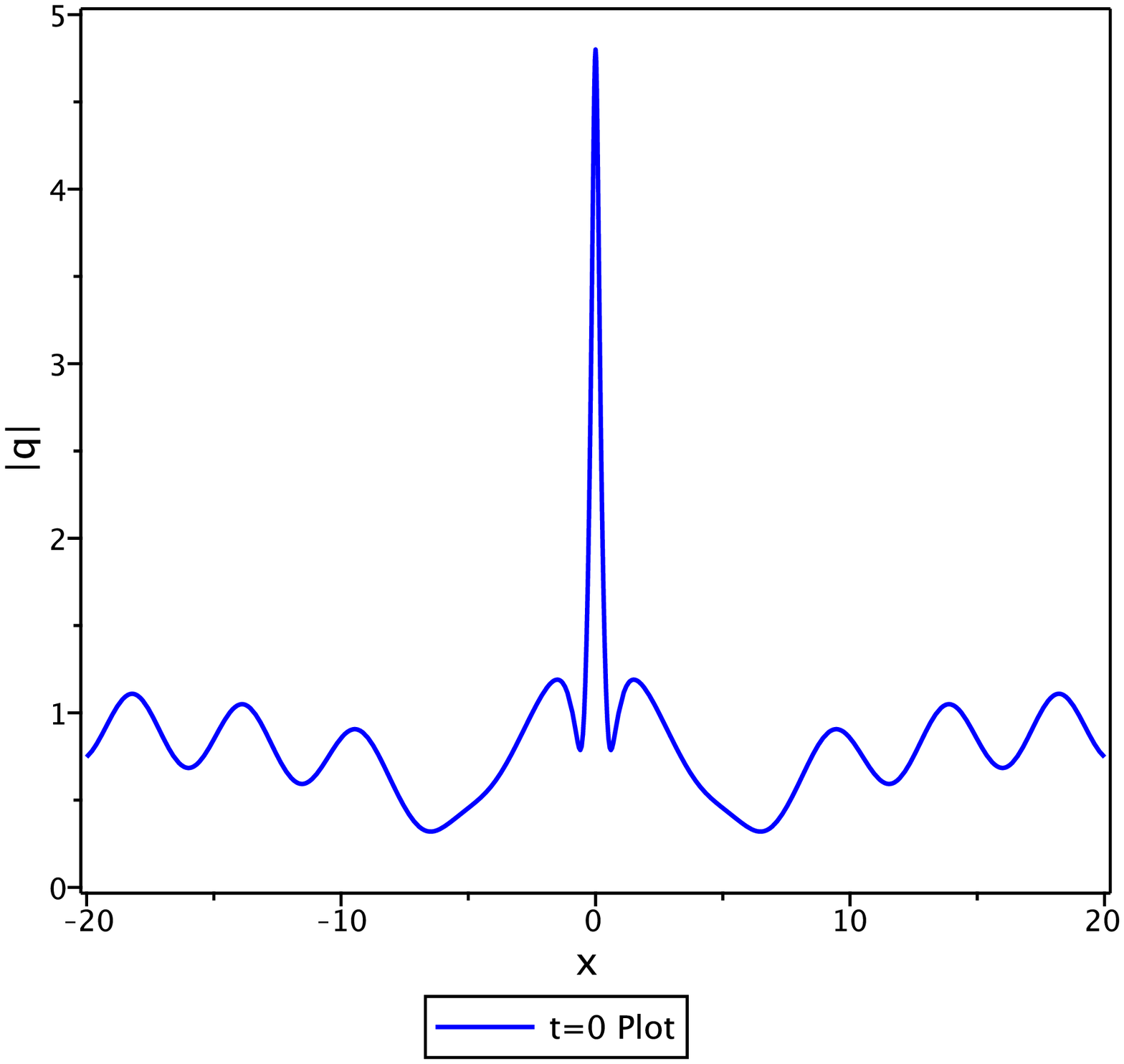}}}\\

$\quad\qquad\textbf{(a)}\qquad\qquad\qquad\qquad\qquad\quad\qquad
 \textbf{(b)}\qquad\qquad\qquad\qquad\qquad\quad\textbf{(c)}$\\
\noindent { \small \textbf{Figure 3.} (Color online) The two rogue periodic wave solution for Eq.\eqref{0.02} with parameter $\beta=-0.1$  and $A=1, a=-1$. $\textbf{(a)}$ Three dimensional plot;
$\textbf{(b)}$ The density plot;
$\textbf{(c)}$ The wave propagation along the $x$-axis at $t=0$.}\\

In general, it is not hard to see that the higher-order rogue periodic wave solution can be generated by Eq.\eqref{6.1}.
However, due to their complex expressions showing these solutions, we just give some plots(see Fig.3) of the two-order rogue periodic wave solution  by taking $N=5$ in Eq.\eqref{6.1}.  In what follows, we would like to use the PINN deep learning scheme to investigate the data-driven periodic wave, one rogue periodic wave, soliton wave and breather wave of the CLL equation \eqref{0.02}.

\section{The data-driven periodic wave,  rogue periodic wave,  soliton wave,   and breather wave}
In the beginning, we focus on the CLL equation \eqref{0.02} along with Dirichlet boundary conditions
\begin{align}\label{16}
\left\{
\begin{array}{lr}
iq_{t}+q_{xx}+iqq^{\ast}q_{x}=0, \quad x\in [x_{0}, x_{1}], t\in [t_{0}, t_{1}],\\
q(x, t_{0})=q_{0}(x),\\
q(x_{0}, t)=q(x_{1},t),
\end{array}
\right.
\end{align}
where $x_{0}, x_{1}$ denote the corresponding boundaries of $x$.  $t_{0}, t_{1}$  are initial and final times of $t$. The $q_{0}(x)$ defines the initial condition. The physics-informed neural networks $f_{u}(x, t)$ and $f_{v}(x, t)$ for the above equation \eqref{16}  can be defined as
\begin{gather}
f_{u}:=-v_{t}+u_{xx}-(u^{2}+v^{2})v_{x},\notag\\
f_{v}:=u_{t}+v_{xx}+(u^{2}+v^{2})u_{x},\label{16.1}
\end{gather}
In terms of the PINN scheme, we can define respectively the complex valued neural network $q(x, t)=u(x, t)+ i v(x, t))$ and $f(x, t)=f_{u}+if_{v}$ into follows by Python:\\
$def \ net_{-}q(self,  \  x,  \  t):$\\
$~~~~~~~$$q \ = \ self.neural_{-}net(tf.concat([x,t],1), \ self.weights, \ self.biases)$\\
$~~~~~~~$$u \ = \ q[:,0:1]$\\
$~~~~~~~$$v \ = \ q[:,1:2]$\\
$~~~~~~~$$return  \  u,  \  v$\\
\\
$def \ net_{-}f(self,  \  x,  \  t):$\\
$~~~~~~~$$u,  \  v  \ =  \  self.net_{-}q(x,t)$\\
$~~~~~~~$$u_{-}t  \ =  \ tf.gradients(u, \  t)[0]$\\
$~~~~~~~$$u_{-}x \ = \ tf.gradients(u, \  x)[0]$\\
$~~~~~~~$$u_{-}xx \  = \ tf.gradients(u_{-}x, \  x)[0]$\\
$~~~~~~~$$v_{-}t \ = \ tf.gradients(v, \  t)[0]$\\
$~~~~~~~$$v_{-}x \ = \ tf.gradients(v, \  x)[0]$\\
$~~~~~~~$$v_{-}xx \ = \ tf.gradients(v_{-}x, \  x)[0]$\\
$~~~~~~~$$f_{-}u \ = \ -v_{-}t \  + \  u_{-}xx \  -  \ (u**2 \  + \  v**2)*v_{-}x$\\
$~~~~~~~$$f_{-}v \ =  \  u_{-}t  \ + \ v_{-}xx \  +  \  (u**2 \  + \  v**2)*u_{-}x$\\
$~~~~~~~$$return f_{-}u, \  f_{-}v$

Next, we will apply the PINN deep learning approach to solve the data-driven   periodic wave, rogue periodic wave, soliton wave, and breather wave  solutions for the CLL equation \eqref{0.02} in detail.

\subsection{The data-driven  periodic wave solution}

Taking $\beta=0.5$ into Eq.\eqref{5} and let $[x_{0}, x_{1}]$ and $[t_{0}, t_{1}]$ in Eq.\eqref{16} as $[-6.0, 6.0]$ and $[0.0, 2.0]$, respectively. We here select the  periodic wave solution at $t=0$ as the initial condition, given by
\begin{align}\label{17}
q(x,0)=q_{0}(x)=\frac{2(\sqrt{3}+1)\cos(\frac{\sqrt{3}}{2}x)}
{(2+\sqrt{3})e^{\frac{(\sqrt{3}-1)ix}{2}}-e^{-\frac{(\sqrt{3}+1)ix}{2}}}.
\end{align}
\\

{\rotatebox{0}{\includegraphics[width=4.5cm,height=4.5cm,angle=0]{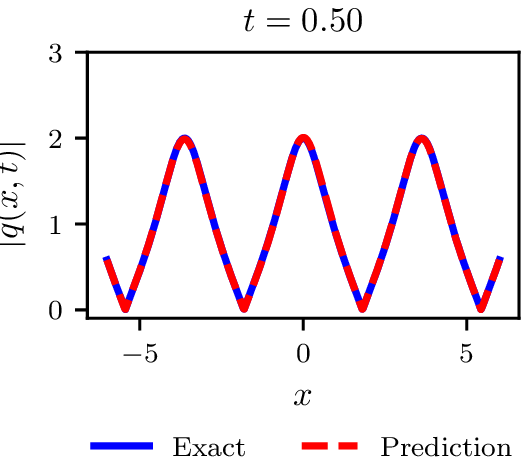}}}
~~~~
{\rotatebox{0}{\includegraphics[width=4.5cm,height=4.5cm,angle=0]{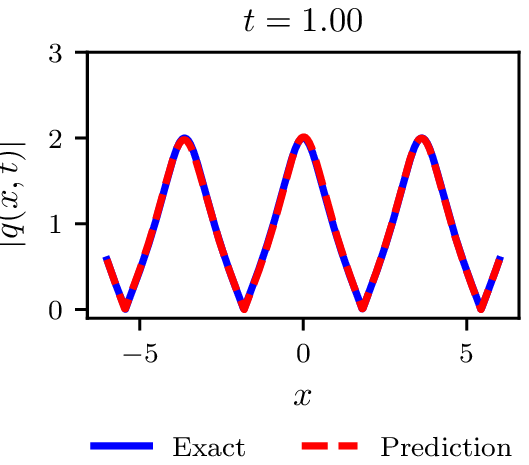}}}
~~~~
{\rotatebox{0}{\includegraphics[width=4.5cm,height=4.5cm,angle=0]{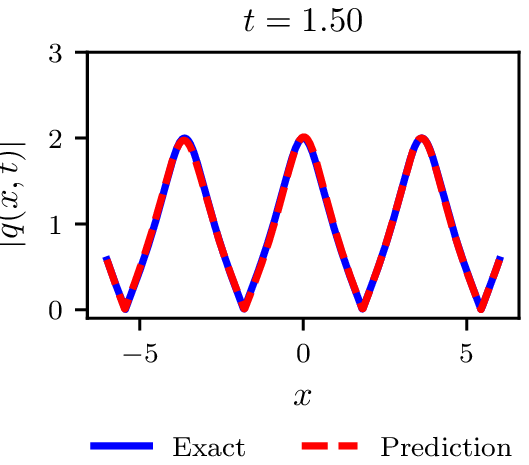}}}\\

$\quad\qquad\quad\quad\qquad\qquad\qquad\quad\quad\qquad\qquad\qquad\textbf{(a)}$\\
{\rotatebox{0}{\includegraphics[width=5.0cm,height=4.0cm,angle=0]{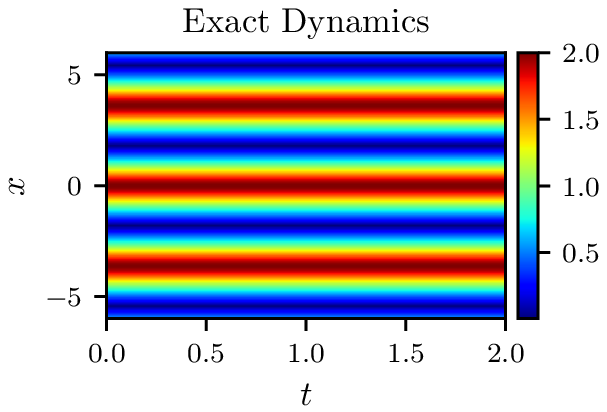}}}
~~~~
{\rotatebox{0}{\includegraphics[width=5.0cm,height=4.0cm,angle=0]{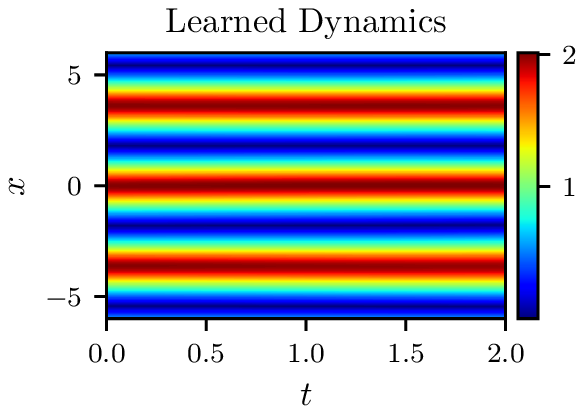}}}
~~~~
{\rotatebox{0}{\includegraphics[width=5.0cm,height=4.0cm,angle=0]{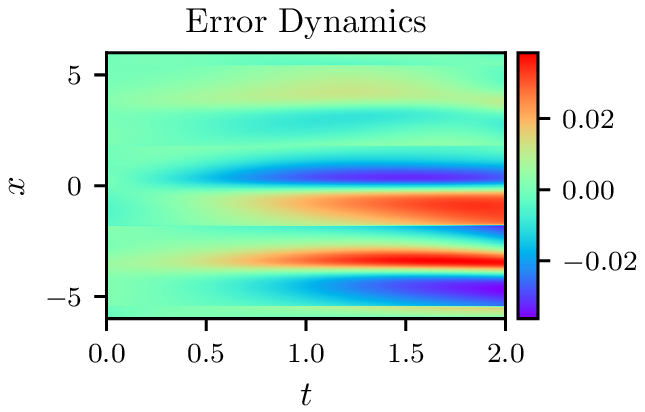}}}\\

$\quad\quad\qquad\qquad\quad\qquad\qquad\quad\quad\qquad\qquad\qquad\textbf{(b)}$\\

{\rotatebox{0}{\includegraphics[width=6.6cm,height=6.0cm,angle=0]{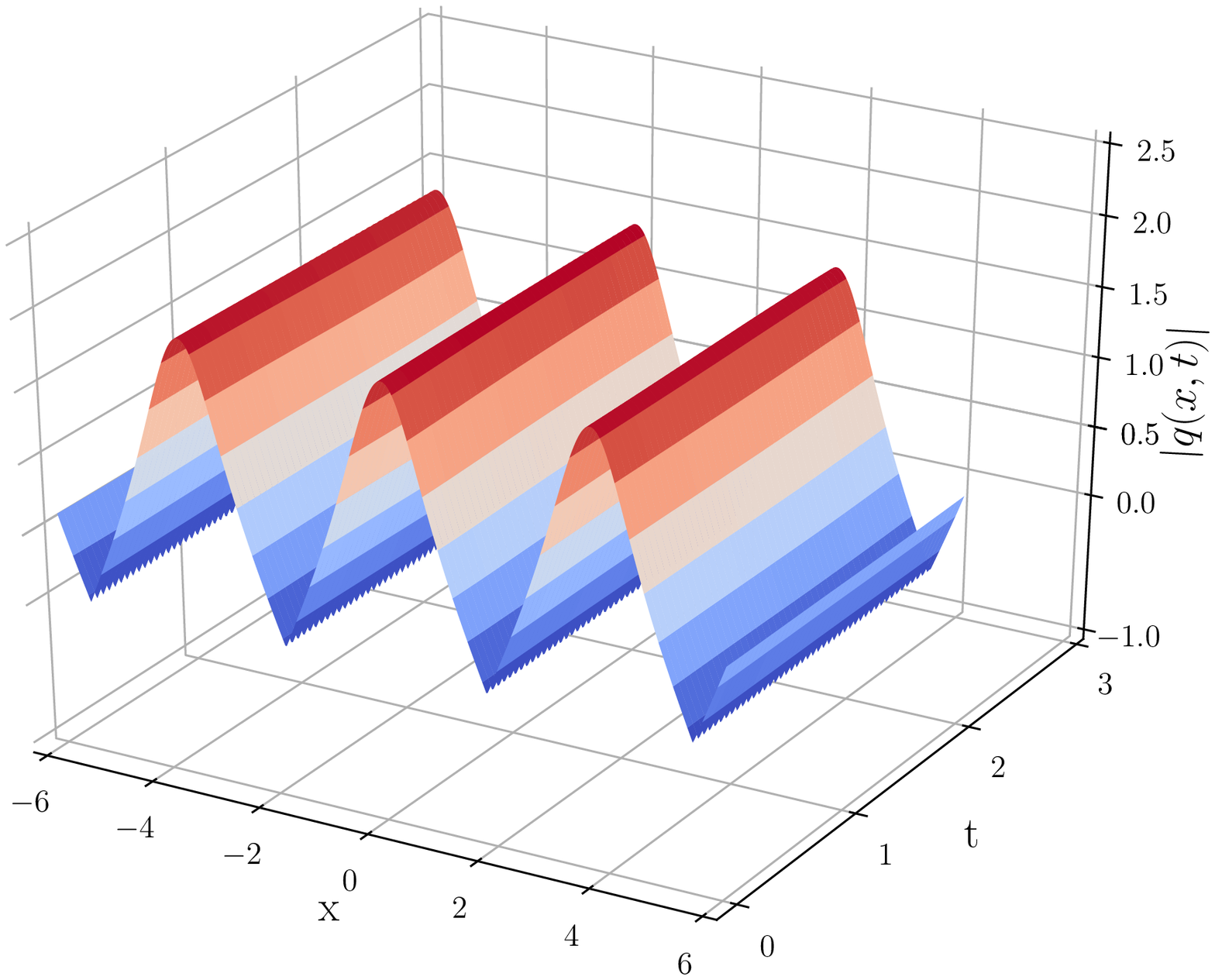}}}
~~~~
{\rotatebox{0}{\includegraphics[width=5.6cm,height=5.0cm,angle=0]{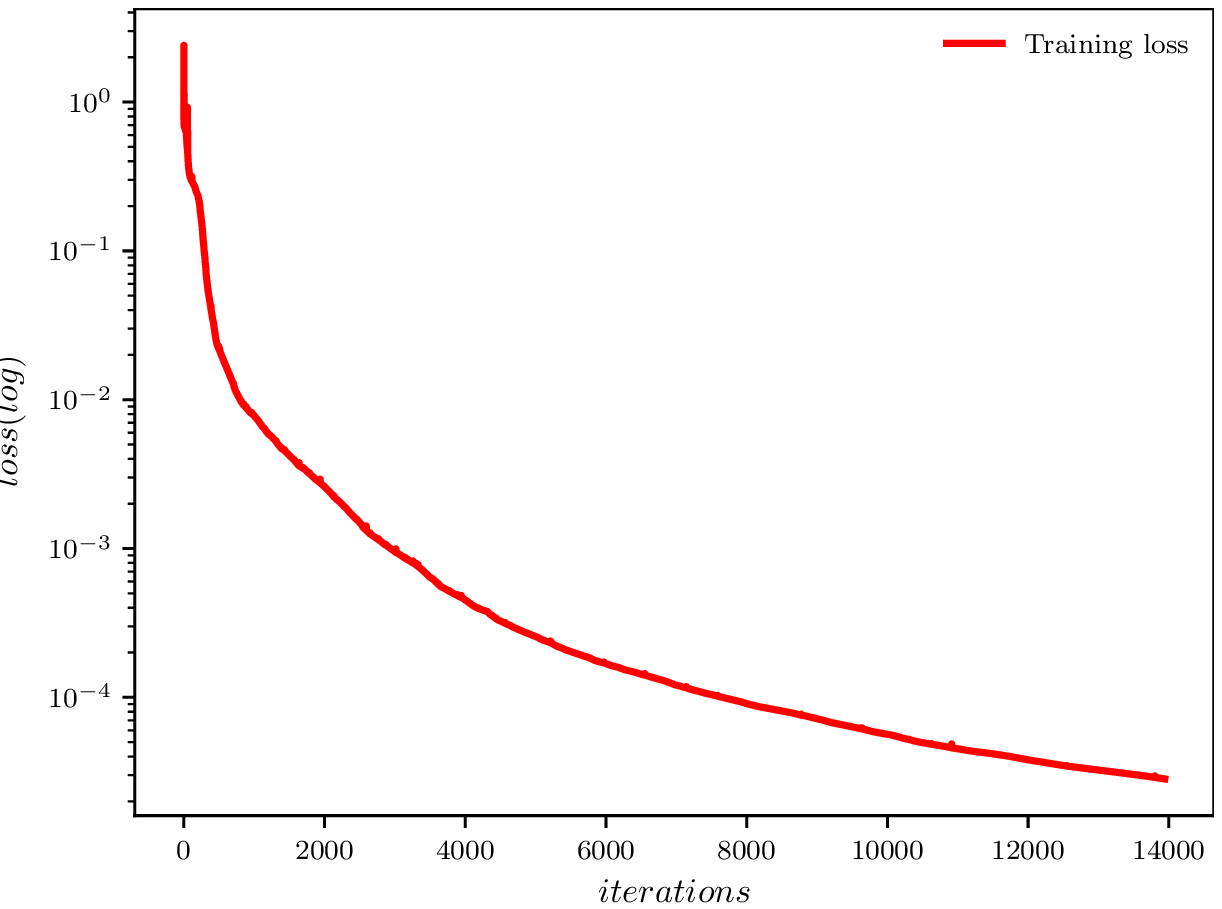}}}\\

$\qquad\qquad\qquad\qquad\textbf{(c)}\qquad\qquad\quad\qquad\qquad\qquad\qquad\qquad\qquad
 \textbf{(d)}$\\
\noindent { \small \textbf{Figure 4.} (Color online) The data-driven periodic wave solution $q(x, t)$ for CLL equation \eqref{0.02}:
$\textbf{(a)}$ The wave propagation plot at three different times;
$\textbf{(b)}$ The density  plot and the  error density diagram;
$\textbf{(c)}$ The three-dimensional plot;
$\textbf{(d)}$ The loss curve figure.}\\

To acquire the original training data, the traditional finite difference method is used to simulate Eq.\eqref{16} with the initial data \eqref{17} by MATLAB. Of which, the spatial region $[-6.0, 6.0]$ is divided into 513 points and time region $[0, 2.0]$ is divided into 401 points. Then, via using the Latin hypercube sampling (LHS) method \cite{PuChen470},  we randomly  extract $N_{q}=100$ from the original initial boundary data and $N_{f}=10000$ collocation points to generate a small training data set containing a subset of the initial boundary. According to obtained training data, using  a 9-hidden-layer deep PINN with 40 neurons per layer, the periodic wave solution $q(x, t)$ is successfully learned by regulating the network parameters and minimize the loss function \eqref{0.6}. The $\mathbb{L}_{2}$-norm error between learning solution and exact solution is 1.141566e-02. The whole learning process takes about 1530.5031 seconds, and iterates 13070 times.

Fig. 4 (a), (b), (c), (d)  display the wave propagation plot at three different times, the density plot, the three-dimensional
motion, the error density diagram, and the loss curve figure,  respectively. From Fig. 4 (a) and (b), we can find that the error between the learning solution and the exact solution is very small.  As shown in Fig. 4(d), the loss curve is quite smooth. These results demonstrate the integrable deep learning method is effective and stable.

\subsection{The data-driven rogue periodic wave solution}
In this part, we devote to research  the data-driven rogue periodic wave solution for CLL equation\eqref{0.02}.
Taking $\beta=-0.02$ into Eq.\eqref{7} and let $[x_{0}, x_{1}]$ and $[t_{0}, t_{1}]$ in Eq.\eqref{16} as $[-12.0, 12.0]$ and $[-1.5, 1.5]$, respectively. We select the rogue periodic wave solution at $t=-1.5$ as the initial condition
\begin{align}\label{18}
q(x,-1.5)=q_{0}(x)=\frac{-11250K_{1}^{+}e^{\frac{3\sqrt{97691}i}{1562500}(1250x-936)-\frac{9i}{8}}
+K_{1}^{-}e^{-\frac{3\sqrt{97691}i}{1562500}(1250x-936)-\frac{9i}{8}}}
{K_{2}^{+}e^{\frac{\sqrt{97691}i}{781250}(1875x-1404)-\frac{ix}{2}}
+11250K_{2}^{-}e^{-\frac{\sqrt{97691}i}{781250}(1875x-1404)-\frac{ix}{2}}},
\end{align}
with
\begin{gather}\label{18.1}
K_{1}^{+}=\sqrt{97691}(\frac{313x^{2}}{1250}
+(\frac{2188i}{1875}-\frac{939}{1250})x+\frac{4321}{3000}-\frac{3i}{2})\notag\\
+\frac{293073x^{2}}{1250}+(\frac{488698i}{1875}-\frac{879219}{1250})x+\frac{12133723}{9000}+\frac{937i}{2},\notag\\
K_{1}^{-}=\sqrt{97691}(-210825x^{2}+(632475+140700i)x+\frac{4848375}{4}-1265625i)\notag\\
-65941425x^{2}+(197824275+43882800i)x-\frac{1516465375}{4}-395578125i,\notag\\
K_{2}^{+}=\sqrt{97691}(211050x^{2}+(140550i-633150)x+\frac{2989875}{2})\notag\\
+65941425x^{2}+(43976550i-197824275)x+\frac{1868465375}{4}-140625i,\notag\\
K_{2}^{-}=\sqrt{97691}(-\frac{x^{2}}{2}+(\frac{3}{2}+\frac{i}{3})x-\frac{77}{24}-\frac{3i}{2})+\frac{938}{9}
-469i+\frac{625ix}{3}.
\end{gather}

{\rotatebox{0}{\includegraphics[width=4.5cm,height=4.5cm,angle=0]{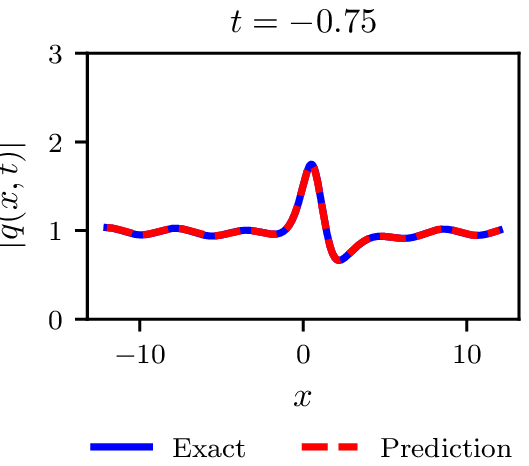}}}
~~~~
{\rotatebox{0}{\includegraphics[width=4.5cm,height=4.5cm,angle=0]{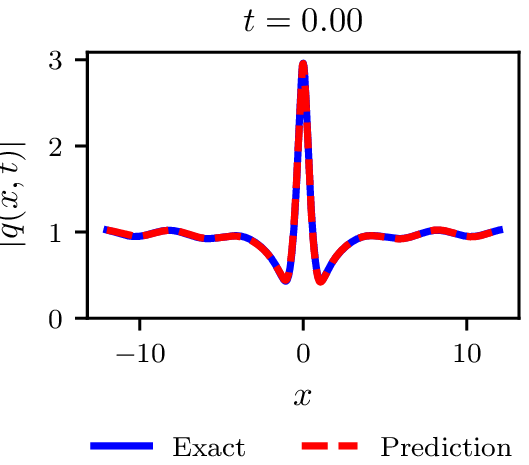}}}
~~~~
{\rotatebox{0}{\includegraphics[width=4.5cm,height=4.5cm,angle=0]{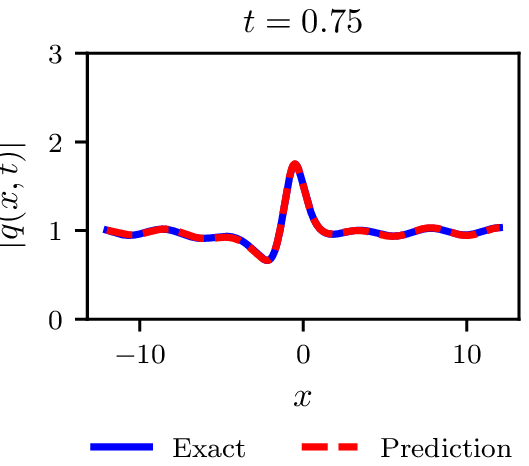}}}\\

$\quad\qquad\quad\quad\qquad\qquad\qquad\quad\quad\qquad\qquad\qquad\textbf{(a)}$\\
{\rotatebox{0}{\includegraphics[width=5.0cm,height=4.0cm,angle=0]{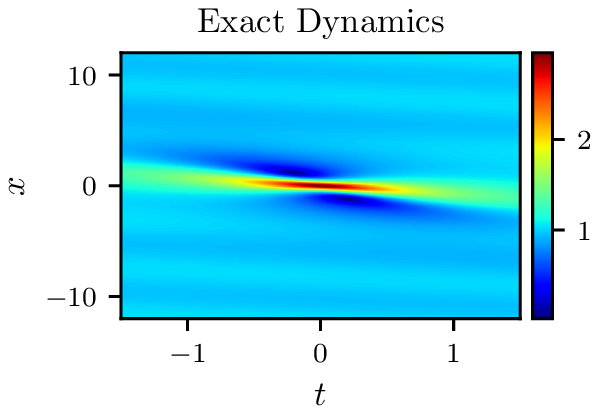}}}
~~~~
{\rotatebox{0}{\includegraphics[width=5.0cm,height=4.0cm,angle=0]{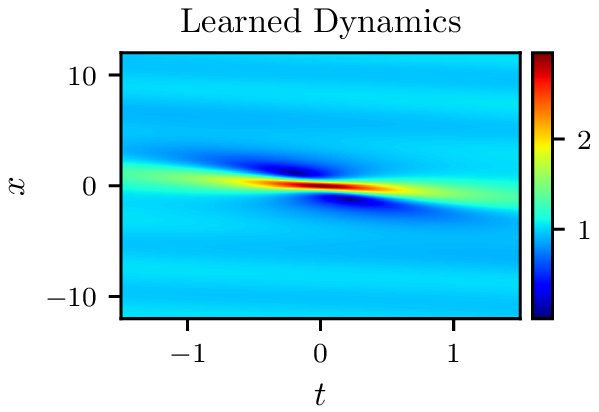}}}
~~~~
{\rotatebox{0}{\includegraphics[width=5.0cm,height=4.0cm,angle=0]{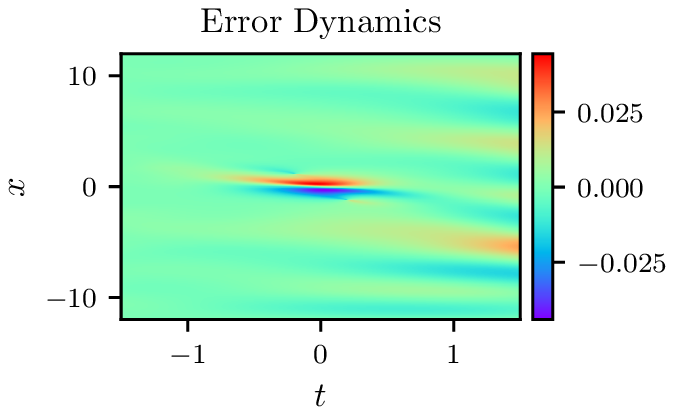}}}\\

$\quad\quad\qquad\qquad\quad\qquad\qquad\quad\quad\qquad\qquad\qquad\textbf{(b)}$\\

{\rotatebox{0}{\includegraphics[width=6.6cm,height=6.0cm,angle=0]{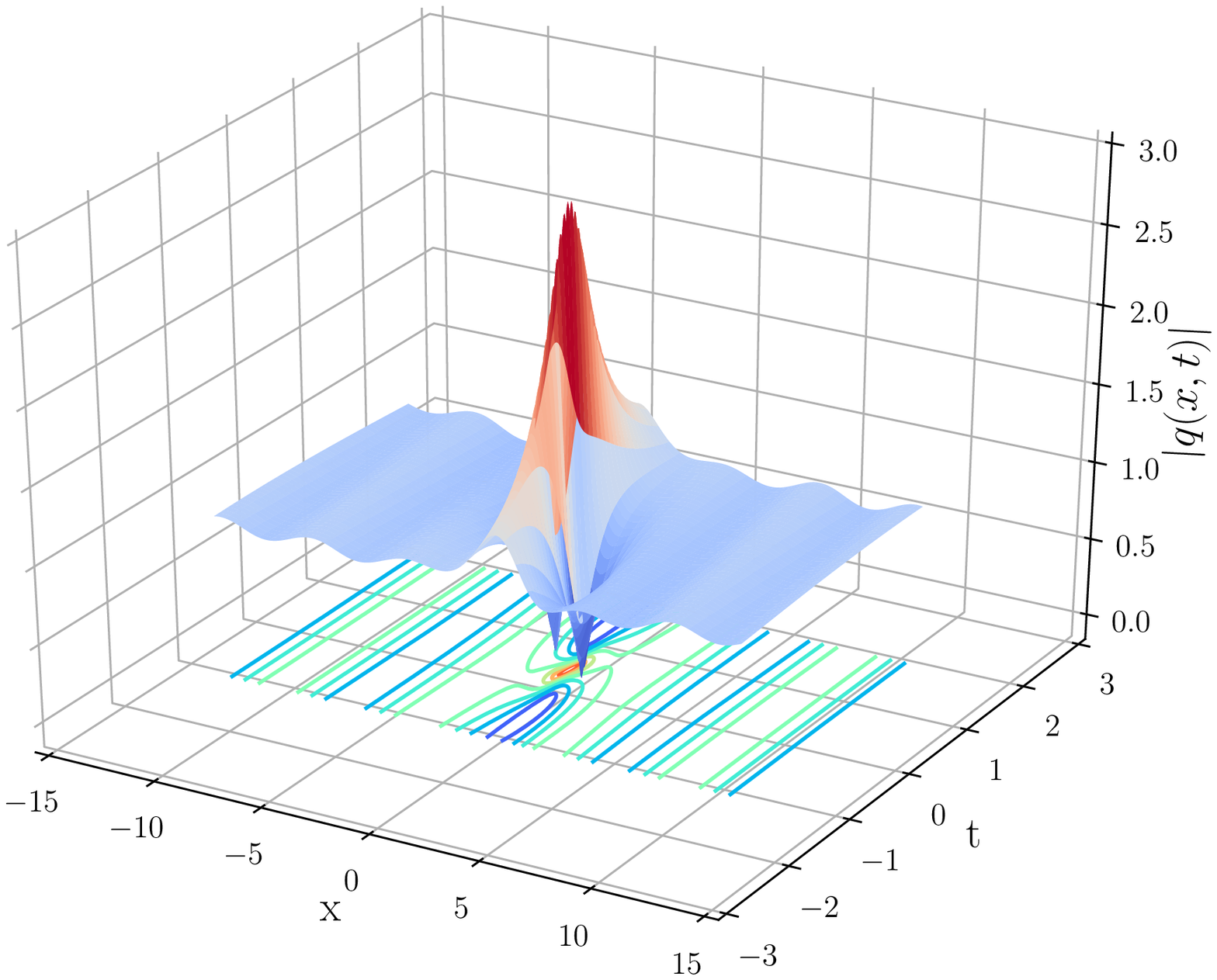}}}
~~~~
{\rotatebox{0}{\includegraphics[width=5.6cm,height=5.0cm,angle=0]{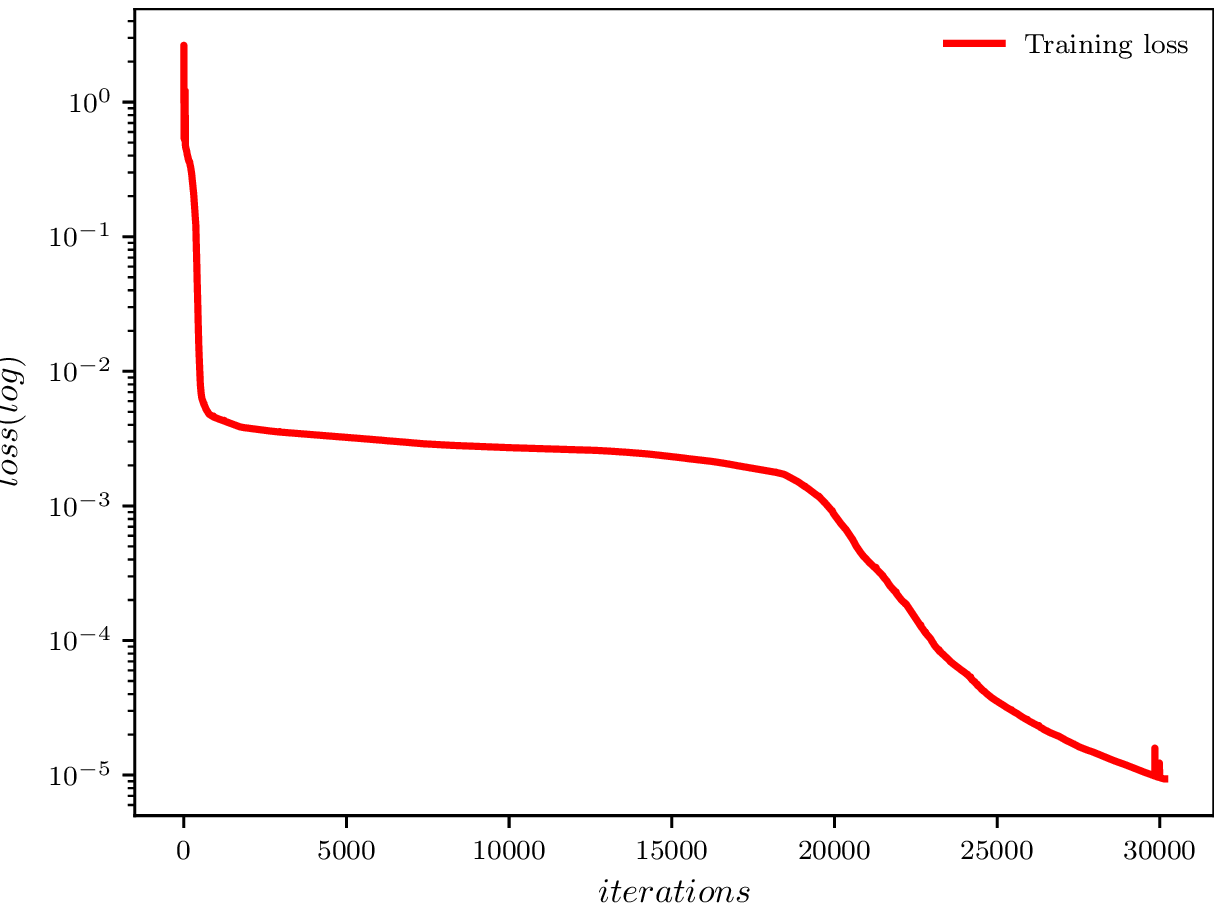}}}\\

$\qquad\qquad\qquad\qquad\textbf{(c)}\qquad\qquad\quad\qquad\qquad\qquad\qquad\qquad\qquad
 \textbf{(d)}$\\
\noindent { \small \textbf{Figure 5.} (Color online) The data-driven rogue periodic wave solution $q(x, t)$ for CLL equation \eqref{0.02}:
$\textbf{(a)}$ The wave propagation plot at three different times;
$\textbf{(b)}$ The density  plot and the  error density diagram;
$\textbf{(c)}$ The three-dimensional plot;
$\textbf{(d)}$ The loss curve figure.}\\

Here, applying the same data discretization method in section 4.1, we generate the initial and boundary value data set with  the spatial region $[-12.0, 12.0]$  dividing into 513 points and temporal region $[-1.5, 1.5]$ into 401 points.  With the help of LHS, a training dataset can be obtained by random sampling $N_{q}=400$ in the original dataset and choosing $N_{f}=10000$ collocation points. Inputting the training dataset into a 9-hidden-layer deep PINN with the first layer is 40 neurons, and the rest is 60 neurons, we successfully generate the learning rogue periodic wave solution which has a $\mathbb{L}_{2}$ error of 6.103918e-03 compared with the exact one. The whole learning process takes about 3551.8307 seconds, and the iteration times is 28208. Due to the rogue periodic wave solution is more multifarious compared with the periodic wave solution, we here choose the bigger sample points and more neurons. However, this does not mean that more neurons are better. When we take 60 neurons per layer, the experiment results are even worse with a  $\mathbb{L}_{2}$ error of 1.560081e-02, training time of 3915.8433 and 31053 number of iterations.

The main results of our experiment are displayed in Fig. 5  including  the wave propagation plot at different time, the  density
plots  for the learning rogue periodic wave solution and exact rogue periodic wave solution, error dynamics diagrams, three dimensional plot and loss curve plot. Through  Fig. 5(a) and (b), we present a comparison between the exact solution and
the learning solution, and it is not hard to find the error is very small. Interestingly, from Fig. 5(d), we can observe that the loss curve is like ``stair'', which does not exist in that one of periodic wave solution.

\subsection{The data-driven  soliton wave solution}
As shown in Ref\cite{He-CLL},  the expression (59) of Ref\cite{He-CLL} will be the bright soliton solution with taking $a=c=1, \beta=0.5$, and be the dark soliton solution with taking $a=c=1, \beta=-0.5$. Let $[x_{0}, x_{1}]$ and $[t_{0}, t_{1}]$ in Eq.\eqref{16} as $[-6.0, 6.0]$ and $[-1.0, 1.0]$ respectively,  the corresponding initial condition for the bright soliton solution is given by
\begin{align}\label{a1}
q(x,-1)=q_{0}(x)=\frac{(1+i)(e^{2+\frac{5i}{4}+(1+\frac{i}{2})x}+e^{\frac{i}{4}(2x+5)})}{ie^{x+2}+1}.
\end{align}
For the dark soliton solution, the initial condition becomes
\begin{align}\label{a2}
q(x,-1)=q_{0}(x)=\frac{-(1+i)(e^{2+\frac{5i}{4}+(1+\frac{i}{2})x}+(1+i)e^{\frac{i}{4}(2x+5)})}{ie^{x+2}-1}.
\end{align}
\\

{\rotatebox{0}{\includegraphics[width=4.5cm,height=4.5cm,angle=0]{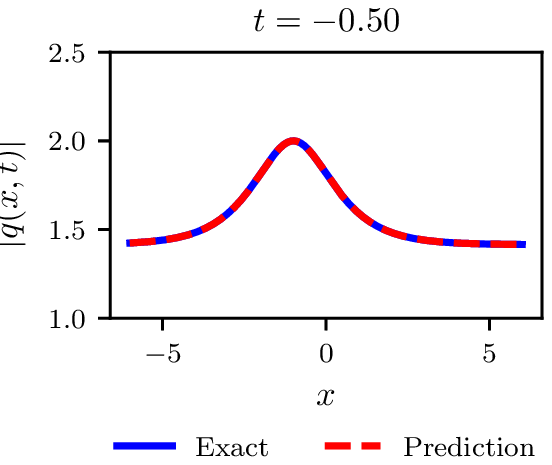}}}
~~~~
{\rotatebox{0}{\includegraphics[width=4.5cm,height=4.5cm,angle=0]{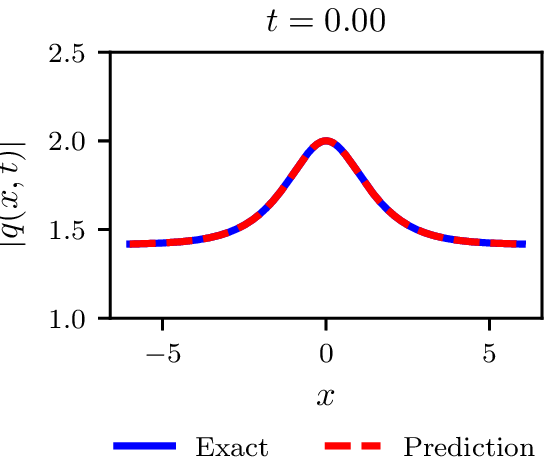}}}
~~~~
{\rotatebox{0}{\includegraphics[width=4.5cm,height=4.5cm,angle=0]{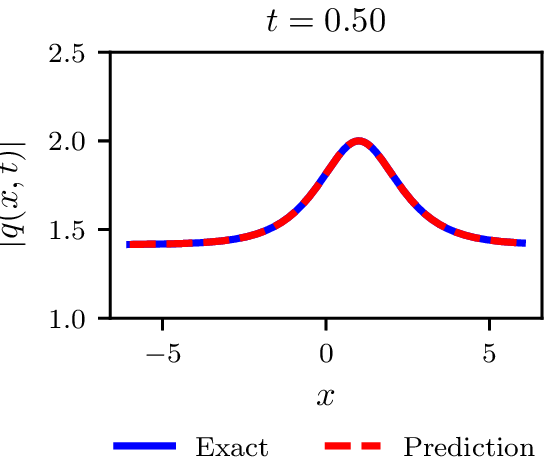}}}\\

$\quad\qquad\quad\quad\qquad\qquad\qquad\quad\quad\qquad\qquad\qquad\textbf{(a)}$\\
{\rotatebox{0}{\includegraphics[width=5.0cm,height=4.0cm,angle=0]{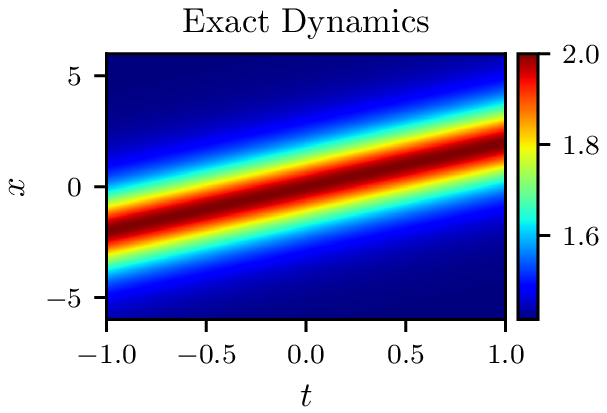}}}
~~~~
{\rotatebox{0}{\includegraphics[width=5.0cm,height=4.0cm,angle=0]{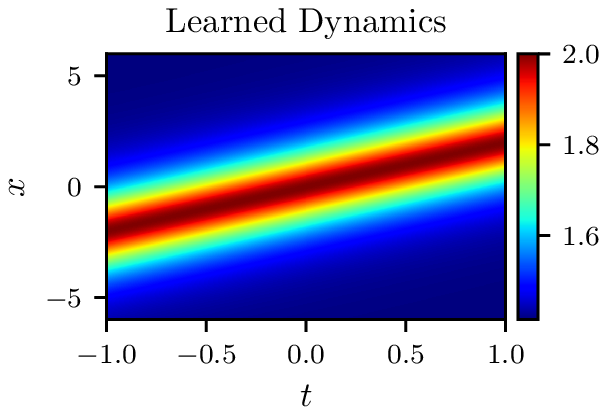}}}
~~~~
{\rotatebox{0}{\includegraphics[width=5.0cm,height=4.0cm,angle=0]{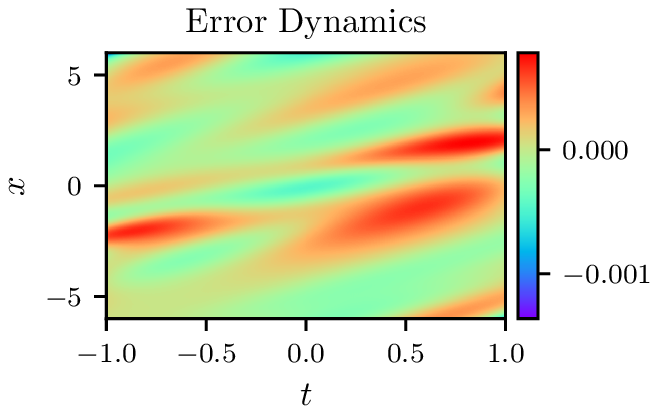}}}\\

$\quad\quad\qquad\qquad\quad\qquad\qquad\quad\quad\qquad\qquad\qquad\textbf{(b)}$\\

{\rotatebox{0}{\includegraphics[width=6.6cm,height=6.0cm,angle=0]{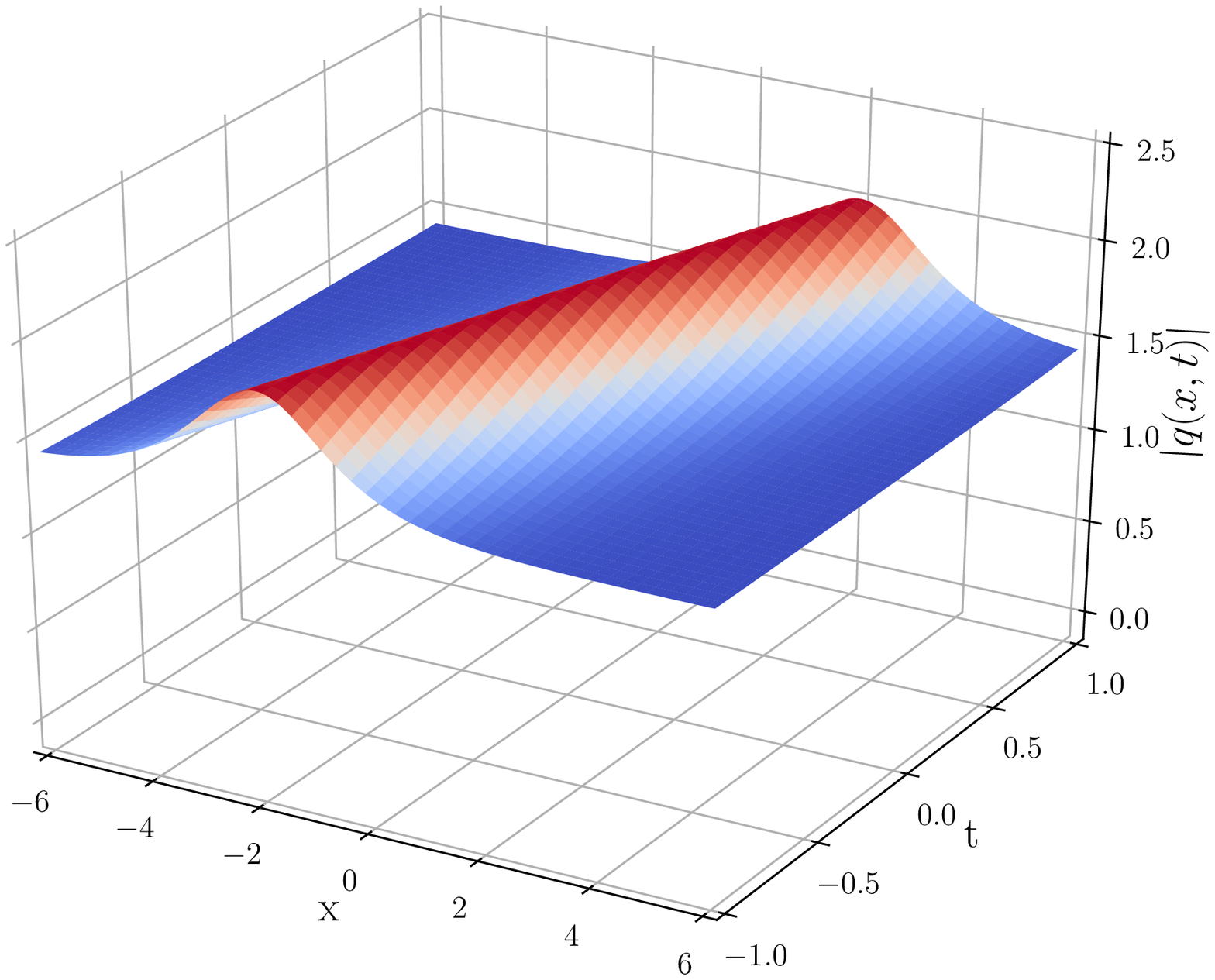}}}
~~~~
{\rotatebox{0}{\includegraphics[width=5.6cm,height=5.0cm,angle=0]{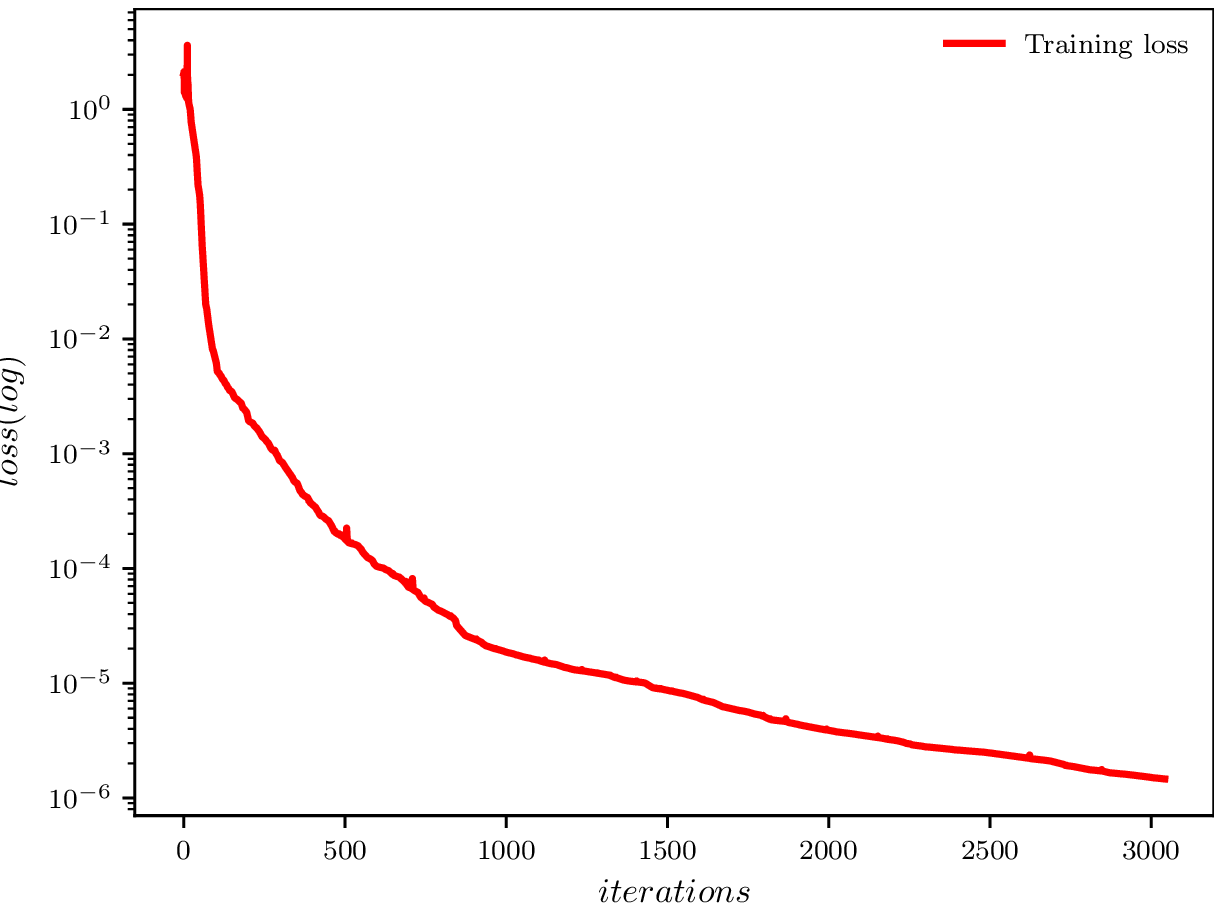}}}\\

$\qquad\qquad\qquad\qquad\textbf{(c)}\qquad\qquad\quad\qquad\qquad\qquad\qquad\qquad\qquad
 \textbf{(d)}$\\
\noindent { \small \textbf{Figure 6.} (Color online) The data-driven bright soliton wave solution $q(x, t)$ for CLL equation \eqref{0.02}:
$\textbf{(a)}$ The wave propagation plot at three different times;
$\textbf{(b)}$ The density  plot and the  error density diagram;
$\textbf{(c)}$ The three-dimensional plot;
$\textbf{(d)}$ The loss curve figure.}\\

Via performing the same data acquisition and training procedures as the section 4.1, it is found, for bright soliton solution, the $\mathbb{L}_{2}$-norm error between learning solution and exact solution is 1.455372e-04, the whole learning process takes about 670.1257 seconds, and iterates 2849 times. For dark soliton solution, the $\mathbb{L}_{2}$-norm error between learning solution and exact solution is 1.924529e-04, the whole learning process takes about 623.2004 seconds, and iterates 2867 times. Fig. 6 and Fig. 7 display the relevant learning outcomes for the bright soliton  and dark soliton,  respectively. According to these experimental results, we find the learning effect for soliton is quite good.

{\rotatebox{0}{\includegraphics[width=4.5cm,height=4.5cm,angle=0]{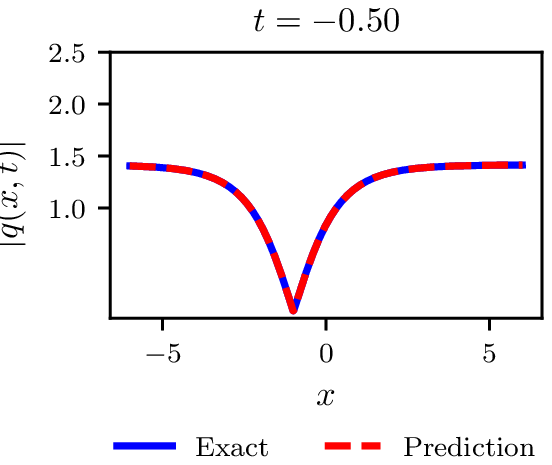}}}
~~~~
{\rotatebox{0}{\includegraphics[width=4.5cm,height=4.5cm,angle=0]{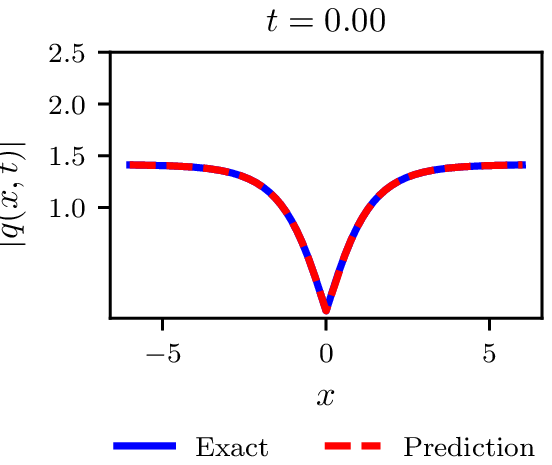}}}
~~~~
{\rotatebox{0}{\includegraphics[width=4.5cm,height=4.5cm,angle=0]{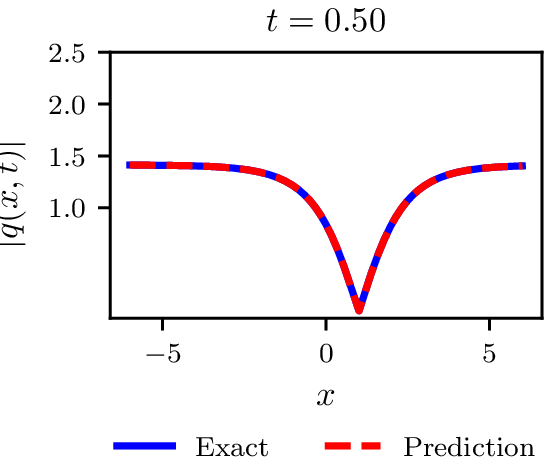}}}\\

$\quad\qquad\quad\quad\qquad\qquad\qquad\quad\quad\qquad\qquad\qquad\textbf{(a)}$\\
{\rotatebox{0}{\includegraphics[width=5.0cm,height=4.0cm,angle=0]{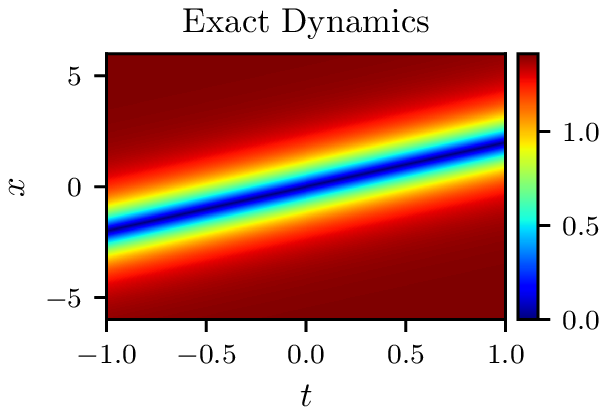}}}
~~~~
{\rotatebox{0}{\includegraphics[width=5.0cm,height=4.0cm,angle=0]{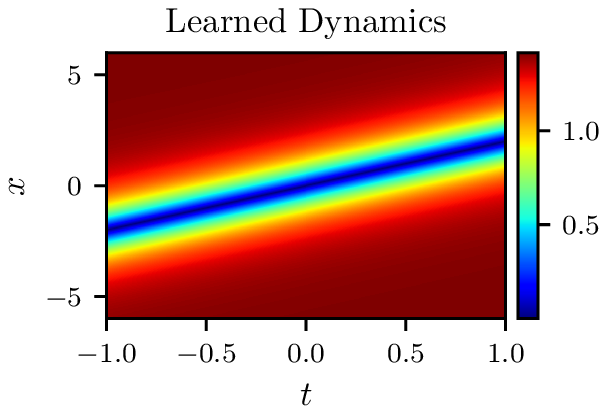}}}
~~~~
{\rotatebox{0}{\includegraphics[width=5.0cm,height=4.0cm,angle=0]{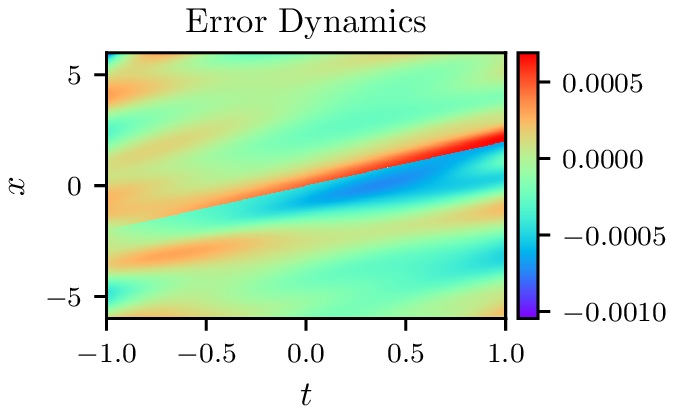}}}\\

$\quad\quad\qquad\qquad\quad\qquad\qquad\quad\quad\qquad\qquad\qquad\textbf{(b)}$\\

{\rotatebox{0}{\includegraphics[width=6.6cm,height=6.0cm,angle=0]{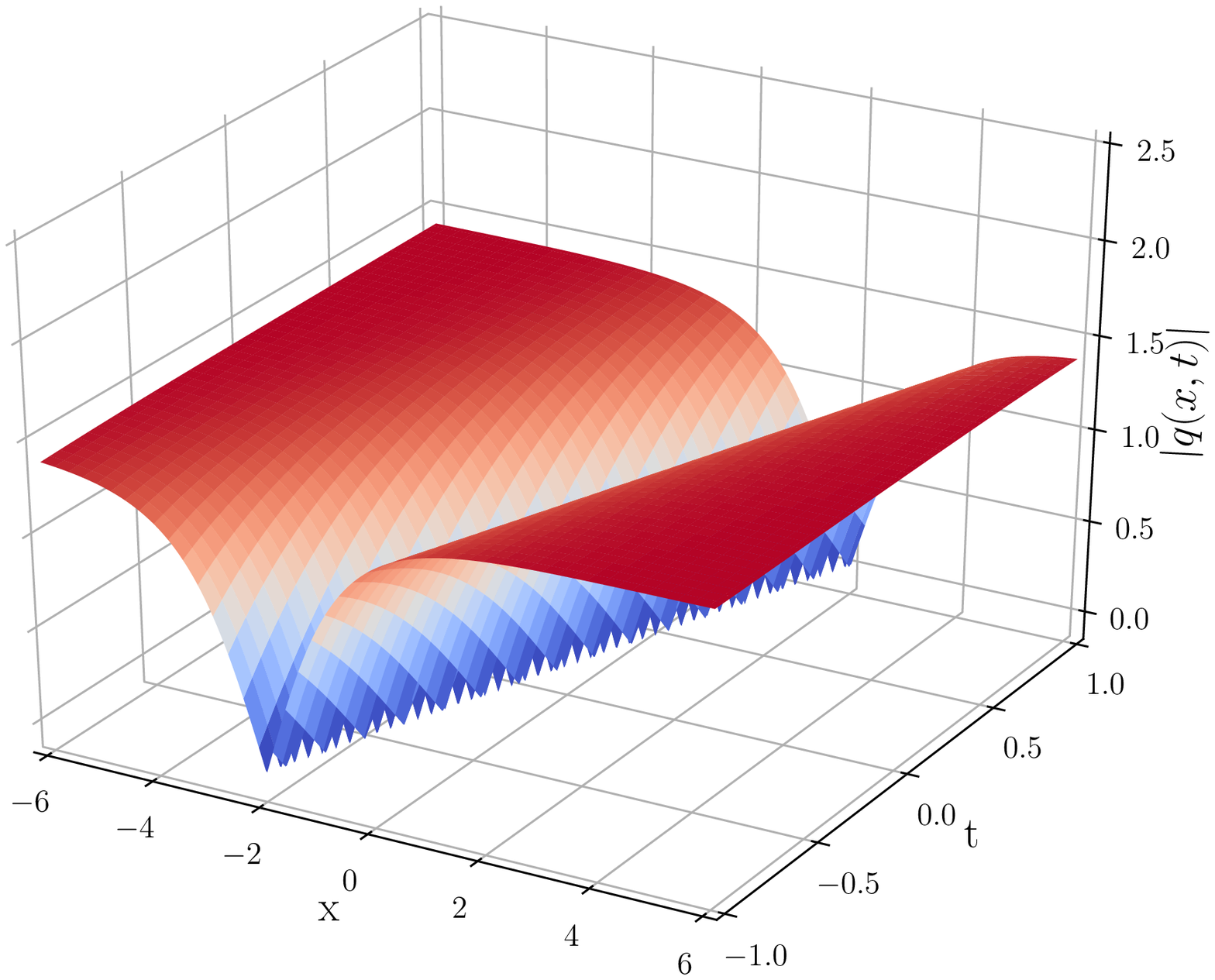}}}
~~~~
{\rotatebox{0}{\includegraphics[width=5.6cm,height=5.0cm,angle=0]{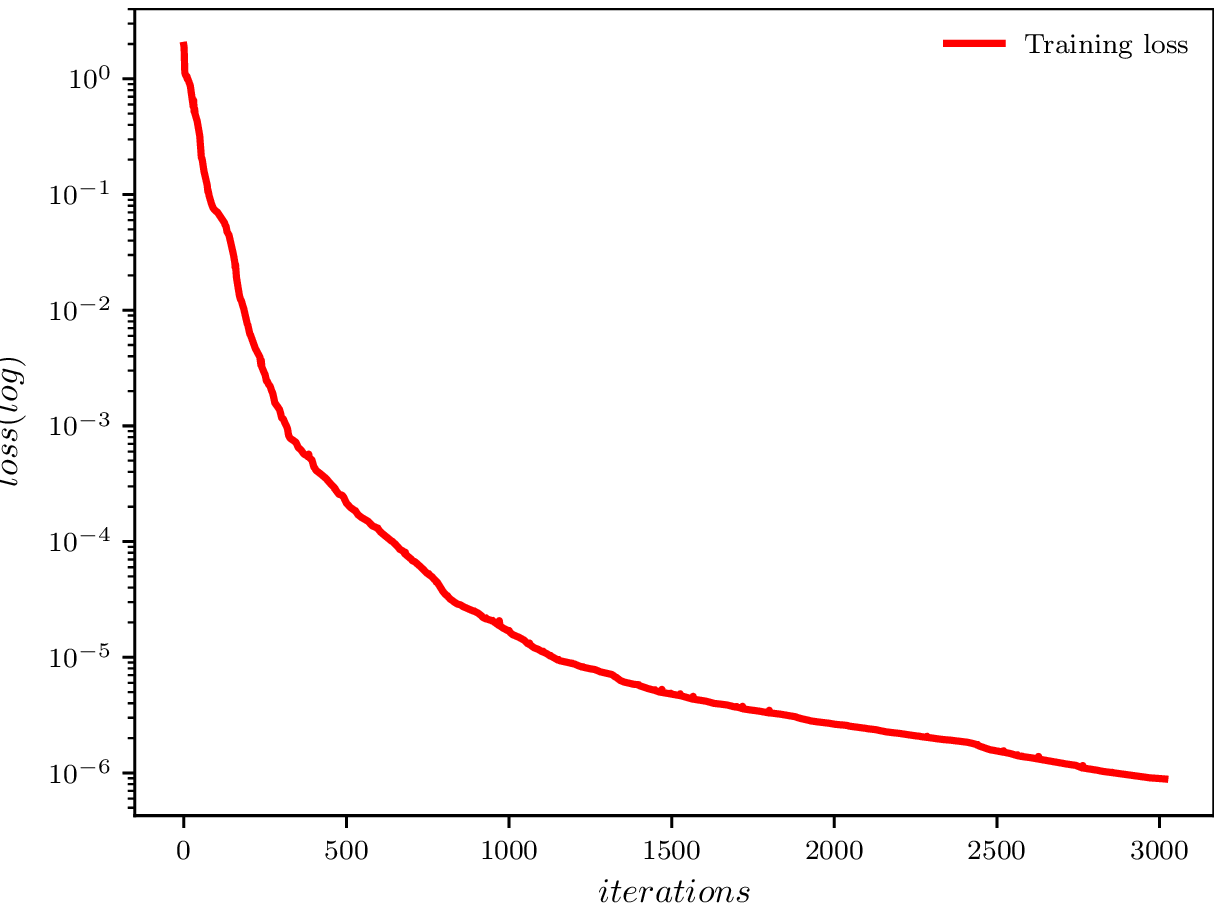}}}\\

$\qquad\qquad\qquad\qquad\textbf{(c)}\qquad\qquad\quad\qquad\qquad\qquad\qquad\qquad\qquad
 \textbf{(d)}$\\
\noindent { \small \textbf{Figure 7.} (Color online) The data-driven dark soliton wave solution $q(x, t)$ for CLL equation \eqref{0.02}:
$\textbf{(a)}$ The wave propagation plot at three different times;
$\textbf{(b)}$ The density  plot and the  error density diagram;
$\textbf{(c)}$ The three-dimensional plot;
$\textbf{(d)}$ The loss curve figure.}\\

\subsection{The data-driven  breather wave solution}
Taking $c=1, \alpha_{1}=0.5, \beta_{1}=0.4,$ into expression (60) in Ref\cite{He-CLL}, and let $[x_{0}, x_{1}]$ and $[t_{0}, t_{1}]$ in Eq.\eqref{16} as $[-12.0, 12.0]$ and $[-3.0, 3.0]$, respectively, and we here select  $t=-3$ as the initial condition for the breather wave solution, given by
\begin{align}\label{b1}
q(x,-1)=q_{0}(x)=\frac{60i(\sqrt{2}-\frac{3}{5})\sin(\frac{3\sqrt{2}(25x-27)}{125})e^{-\frac{17i}{625}(25x+24)}
-H_{1}e^{-\frac{17i}{625}(25x+24)}}
{H_{2}},
\end{align}
where
\begin{gather}\label{b2}
H_{1}=180\sqrt{2}i\sinh(\frac{36\sqrt{2}}{25})-108i\sinh(\frac{36\sqrt{2}}{25})
-21\sqrt{2}\cosh(\frac{36\sqrt{2}}{25})\notag\\
+60\sqrt{2}\cos(\frac{3\sqrt{2}(25x-27)}{125})
-200\cos(\frac{3\sqrt{2}(25x-27)}{125})+70\cosh(\frac{36\sqrt{2}}{25}),\notag\\
H_{2}=(60\sqrt{2}-200)\cos(\frac{3\sqrt{2}(25x-27)}{125})+(60\sqrt{2}i-36i)\sin(\frac{3\sqrt{2}(25x-27)}{125})\notag\\
+(250-75\sqrt{2})\cosh(\frac{36\sqrt{2}}{25})+60(\sqrt{2}-\frac{3}{5})i\sinh(\frac{36\sqrt{2}}{25}).
\end{gather}
\\

{\rotatebox{0}{\includegraphics[width=4.5cm,height=4.5cm,angle=0]{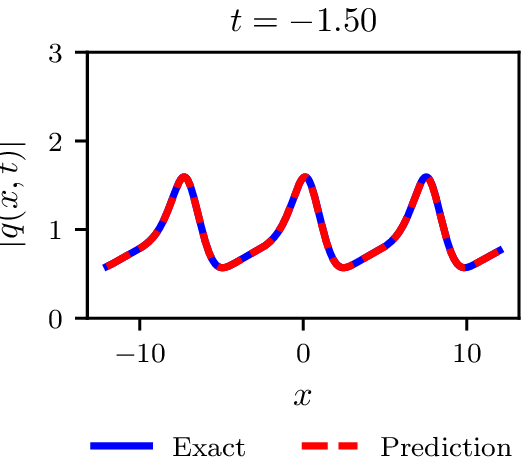}}}
~~~~
{\rotatebox{0}{\includegraphics[width=4.5cm,height=4.5cm,angle=0]{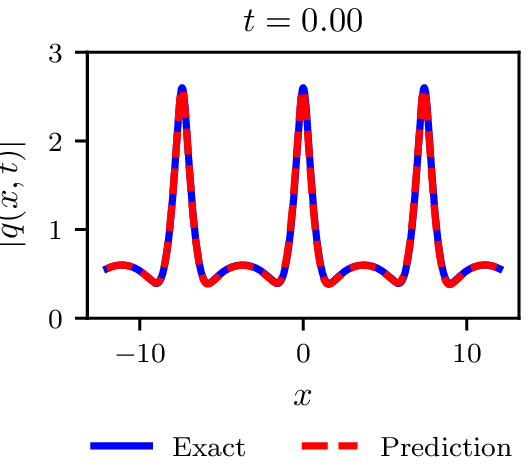}}}
~~~~
{\rotatebox{0}{\includegraphics[width=4.5cm,height=4.5cm,angle=0]{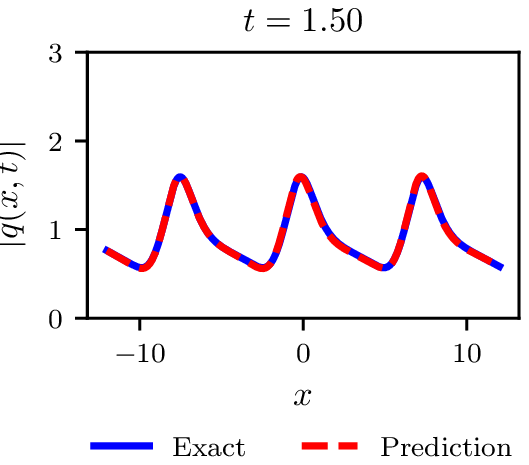}}}\\

$\quad\qquad\quad\quad\qquad\qquad\qquad\quad\quad\qquad\qquad\qquad\textbf{(a)}$\\
{\rotatebox{0}{\includegraphics[width=5.0cm,height=4.0cm,angle=0]{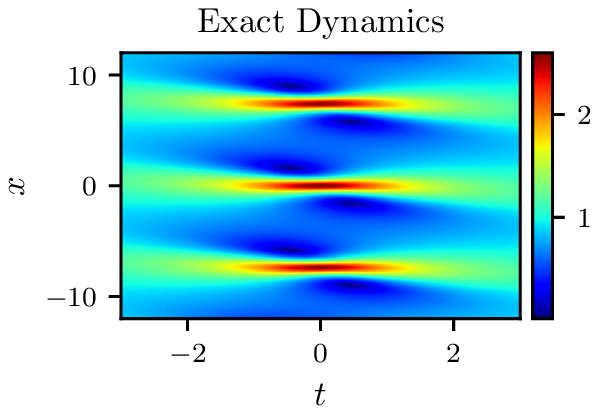}}}
~~~~
{\rotatebox{0}{\includegraphics[width=5.0cm,height=4.0cm,angle=0]{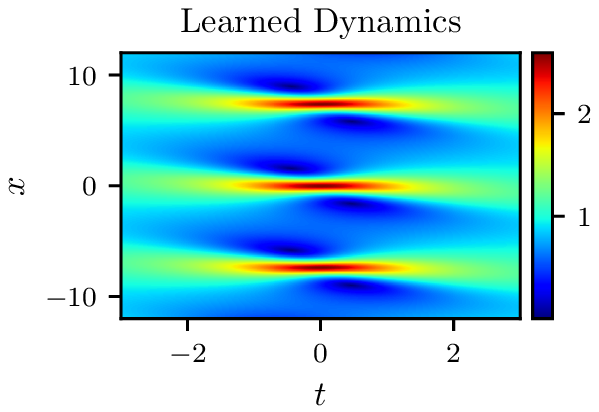}}}
~~~~
{\rotatebox{0}{\includegraphics[width=5.0cm,height=4.0cm,angle=0]{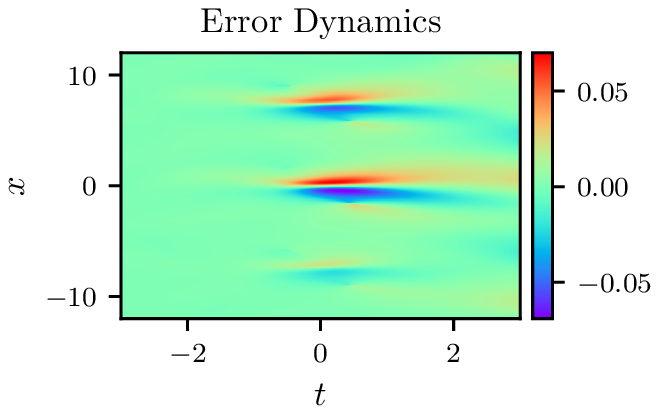}}}\\

$\quad\quad\qquad\qquad\quad\qquad\qquad\quad\quad\qquad\qquad\qquad\textbf{(b)}$\\

{\rotatebox{0}{\includegraphics[width=6.6cm,height=6.0cm,angle=0]{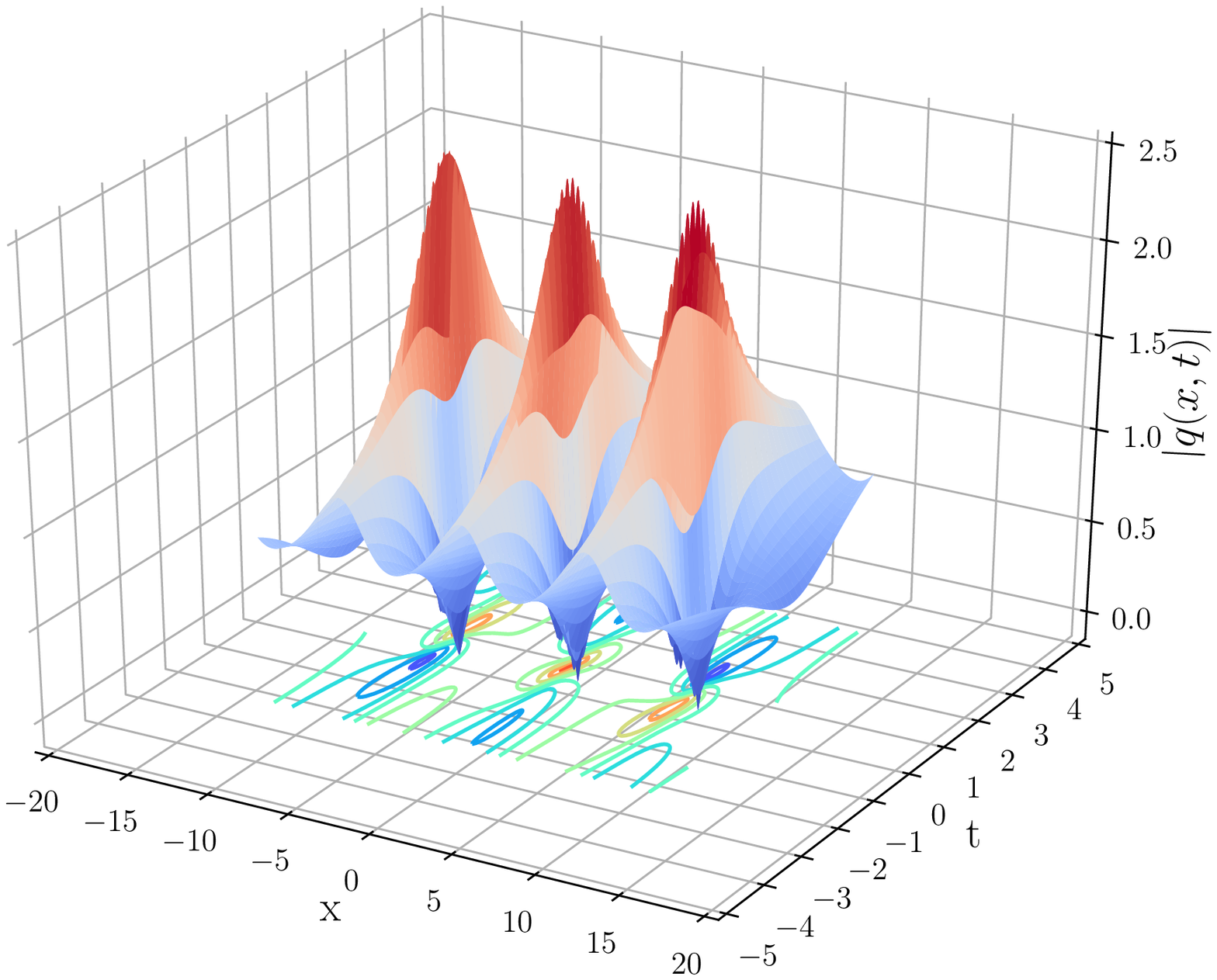}}}
~~~~
{\rotatebox{0}{\includegraphics[width=5.6cm,height=5.0cm,angle=0]{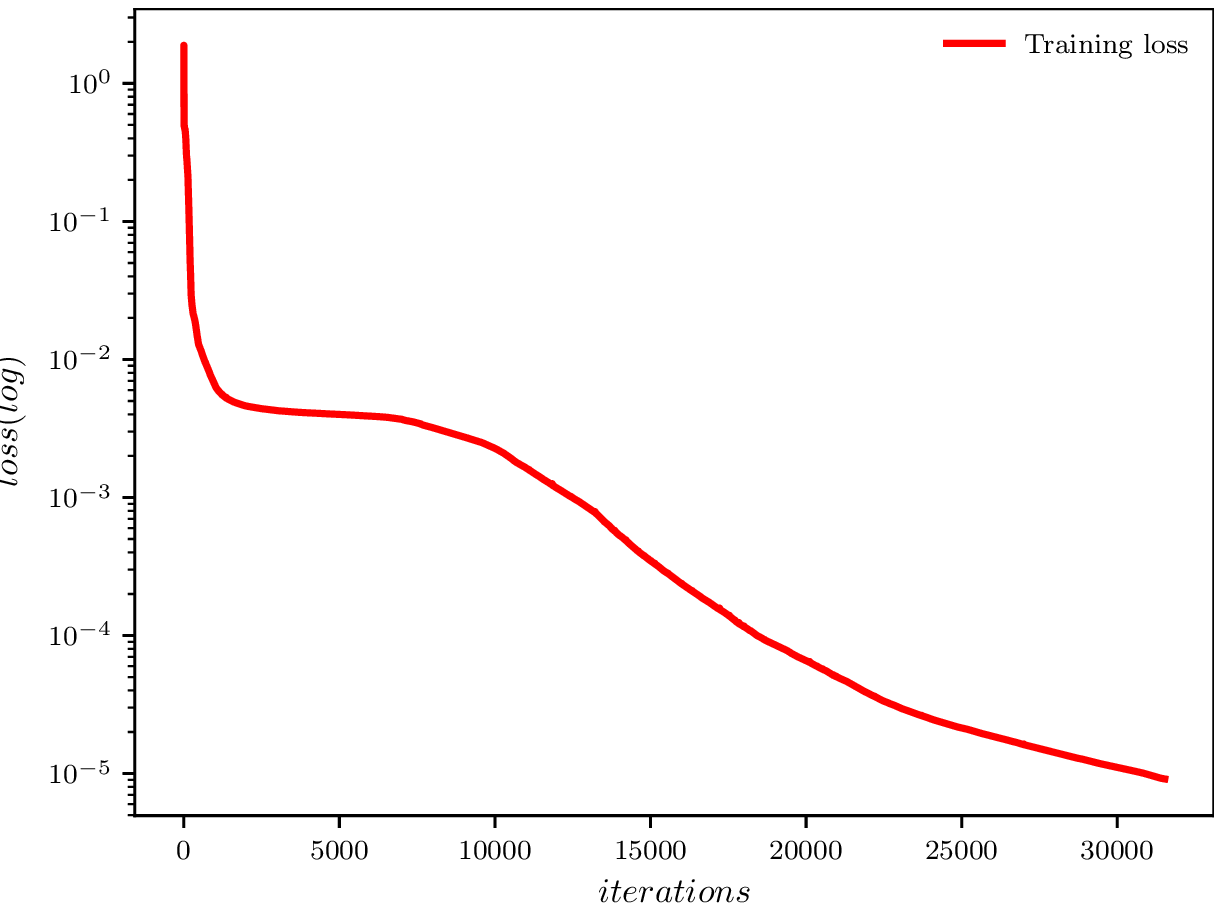}}}\\

$\qquad\qquad\qquad\qquad\textbf{(c)}\qquad\qquad\quad\qquad\qquad\qquad\qquad\qquad\qquad
 \textbf{(d)}$\\
\noindent { \small \textbf{Figure 8.} (Color online) The data-driven breather wave solution $q(x, t)$ for CLL equation \eqref{0.02}:
$\textbf{(a)}$ The wave propagation plot at three different times;
$\textbf{(b)}$ The density  plot and the  error density diagram;
$\textbf{(c)}$ The three-dimensional plot;
$\textbf{(d)}$ The loss curve figure.}\\

Using the same data discretization method as in section 4.1, we obtain the initial and boundary value data set with  the spatial region $[-12.0, 12.0]$  dividing into 513 points and temporal region $[-3.0, 3.0]$ into 401 points. Being different from the case of section 4.1, we take the $N_{q}=400$  boundary sample point and  $N_{f}=20000$ collocation points. Besides, a 9-hidden-layer deep PINN with 60 neurons per layer is choosed here. After training, the neural network model reaches a $\mathbb{L}_{2}$ error of 1.156422e-02 compared with the exact one. The whole learning process takes about 4212.7363 seconds, and the iteration times is 29376.
Fig. 8 presents  the relevant dynamical behaviors and error analysis for the breather wave solution.  Being analogous to the loss curve of rogue periodic wave solution, there is a gentle interregion for the loss curve in Fig. 8 (d).

\section{Conclusion}
In this paper, we have applied the odd-th order DT to derive the exact periodic wave and
rogue periodic wave for CLL equation\eqref{0.02}. Then, in terms of the obtained exact solutions, PINN deep learning method  was  introduced  to solve the periodic wave and rogue periodic wave involving the CLL equation\eqref{0.02}. It is worth mentioning that the deep learning for the rogue periodic wave is first realized to solve the partial differential equation.  As well as, we applied the PINN deep learning
approach to solve the data-driven soliton wave and breather wave solutions for CLL equation\eqref{0.02}. Our results indicate that the errors between the exact solutions with the ones  generated by PINN deep learning method  is vary  small, which verifies the integrable deep learning method is effective and stable.  Compared with the traditional numerical methods, the PINN deep learning method has no grid size limitation. In addition, due to physical constraints, the network is trained with only a small amount of data and has better physical interpretation.  This method opens up a new way to solve the integrabel  and unintegrabel systems by using deep learning and find some novel  models in the interdisciplinary field of applied mathematics and computational science.
Remarkably, by selecting a certain time domain, PINN method has a good training effect. However, with a wider range
of time interval, the training effect will not be as good as we expected. Especially for rogue periodic wave, the effect is only good in a small time range. Therefore, in the future, we will solve the problem about how to simulate the rogue periodic wave well in a large spatio-temporal scale, such as using a reservoir computing approach, or selecting the Lax equation as the physical constraints rather than the equation itself.


\end{document}